\documentclass[11pt,preprint2]{aastex}

\newcommand{\Msun}{\hbox{$\hbox{M}_\odot\;$}}
\newcommand{\Rsun}{\hbox{$\hbox{R}_\odot\;$}}
\newcommand{\kms}{\hbox{${\rm km}\:{\rm s}^{-1}\;$}}
\newcommand{\Msuno}{\hbox{$\hbox{M}_\odot$}}
\newcommand{\Rsuno}{\hbox{$\hbox{R}_\odot$}}
\newcommand{\kmso}{\hbox{${\rm km}\:{\rm s}^{-1}$}}
\newcommand{\teff}{$T_{\rm eff}\;$}  
\newcommand{\teffo}{$T_{\rm eff}$}  
  
\newcommand{\loggo}{$\log\;g$}

\shorttitle{Stellar Abundances in XTE J1118+480}
\shortauthors{J. I. Gonz\'alez Hern\'andez et al.}

\begin{document}


\title{Chemical Abundances of the Secondary Star \\ in the Black Hole
X-ray Binary XTE J1118+480}


\author{Jonay I. Gonz\'alez Hern\'andez\altaffilmark{1,2,3}, Rafael
Rebolo\altaffilmark{3,4}, Garik Israelian\altaffilmark{3},
Alexei V. Filippenko\altaffilmark{5}, Ryan Chornock\altaffilmark{5},
Nozomu Tominaga\altaffilmark{6}, Hideyuki Umeda\altaffilmark{6}, and 
Ken'ichi Nomoto\altaffilmark{6,7,8}}


\altaffiltext{1}{Observatoire de Paris-Meudon, GEPI, 5 place Jules
Janssen, 92195 Meudon Cedex, France; Jonay.Gonzalez-Hernandez@obspm.fr}
\altaffiltext{2}{CIFIST Marie Curie Excellence Team}
\altaffiltext{3}{Instituto de Astrof{\'\i }sica de Canarias, E-38205 La Laguna,
Tenerife, Spain; rrl@iac.es, gil@iac.es}
\altaffiltext{4}{Consejo Superior de Investigaciones 
Cient{\'\i}ficas, Spain.}
\altaffiltext{5}{Department of Astronomy, Uni\-ver\-si\-ty of
California, Berkeley, CA 94720-3411, USA; \\ alex@astro.berkeley.edu,
rchornock@astro.berkeley.edu} 
\altaffiltext{6}{Department of Astronomy, School of Science, University of
Tokyo, Bunkyo-ku, Tokyo 113-0033, Japan;
tominaga@astron.s.u-tokyo.ac.jp, umeda@astron.s.u-tokyo.ac.jp, 
nomoto@astron.s.u-tokyo.ac.jp}
\altaffiltext{7}{Research Center for the Early Universe, School of
Science, University of Tokyo, Bunkyo-ku, Tokyo 113-0033, Japan.} 
\altaffiltext{8}{Institute for the Physics and Mathematics of the
Universe, University of Tokyo, Kashiwa, Chiba 277-8582, Japan.} 


\begin{abstract}

Following the recent abundance measurements of Mg, Al, Ca, Fe, and Ni
in the black hole X-ray binary \mbox{XTE J1118+480} using
medium-resolution Keck~II/ESI spectra of the secondary star
(Gonz\'alez Hern\'andez et al. 2006), we perform a detailed abundance
analysis including the abundances of Si and Ti. These element
abundances, higher than solar, indicate that the black hole in this
system formed in a supernova event, whose nucleosynthetic products
could pollute the atmosphere of the secondary star, providing clues on
the possible formation region of the system, either Galactic halo,
thick disk, or thin disk. We explore a grid of explosion models
with different He core masses, metallicities, and geometries.
Metal-poor models associated with a formation scenario in the Galactic
halo provide unacceptable fits to the observed abundances, allowing
us to reject a halo origin for this X-ray binary. The thick-disk
scenario produces better fits, although they require substantial
fallback and very efficient mixing processes between the inner layers
of the explosion and the ejecta, making quite unlikely an origin
in the thick disk. The best agreement between the model predictions
and the observed abundances is obtained for metal-rich progenitor
models. In particular, non-spherically symmetric models are able to
explain, without strong assumptions of extensive fallback and mixing,
the observed abundances. Moreover, asymmetric mass ejection in
a supernova explosion could account for the required impulse necessary
to launch the system from its formation region in the Galactic thin
disk to its current halo orbit.
 
\end{abstract}


\section{Introduction}
       
The low-mass X-ray binary \mbox{XTE J1118+480} is the first identified
black hole moving in Galactic halo regions (Wagner et al. 2001;
Mirabel et al. 2001). Since it was discovered during a faint outburst
on UT 2000 March 29 (Remillard et al.\ 2000), it has been intensively
studied in both the X-ray and optical spectral regions. During the
decay of the outburst, McClintock et al. (2001) and Wagner et al.
(2001) determined the radial-velocity curve of the companion star,
yielding a mass function $f(M) \approx 6$~\Msuno. The companion star
was classified as a late-type main-sequence star with a mass of
0.1--0.5~{\Msuno} (Wagner et al.\ 2001). 

By modelling the light curve, McClintock et al. (2001) 
derived a lower limit to the
orbital inclination, $i \ga 55^\circ$, and consequently an upper limit
to the black hole mass of $M_{\rm BH} \la 10$~\Msuno. Additional
evidence for a high inclination ($i \ga 60^\circ$) comes from
measurements of tidal distortion (Frontera et al. 2001), whereas the
lack of dips or eclipses for a Roche-lobe filling secondary yields
upper limits of $i \ga 80^\circ$ and $M_{\rm BH} \ga 7.1$~\Msuno. 
Later, Gelino et al. (2006) derived an orbital inclination of
$68^\circ\;\pm\;2^\circ$, by modeling the optical and infrared
ellipsoidal light curves of the system in quiescence. This value of
the inclination allowed them to better constrain the black hole mass
at $M_{\rm BH} = 8.53 \pm 0.60$~\Msuno.   

The system is placed in the Galactic halo, with an extraordinarily
high Galactic latitude ($b \approx 62.3^{\circ}$), and a height of
$\sim$1.6 kpc above the Galactic plane, according to its distance of
$1.85\pm0.36$ kpc (Wagner et al.\ 2001). This appears surprising since
all other black hole binaries are located in the Galactic disk. 
An accurate measurement of
its proper motion coupled with its distance provides space-velocity
components ($U$, $V$) which seem consistent with those of some old halo
globular clusters (Mirabel et al.\ 2001). This opened the possibility
that the system originated in the Galactic halo, and therefore, that the
black hole could be either the remnant of a supernova (SN) in the very
early Galaxy or the result of direct collapse of an ancient massive star.
However, the galactocentric orbit crossed the Galactic plane many
times in the past, and the system could have formed in the
Galactic disk and been launched into its present orbit as a
consequence of the ``kick'' imparted during the SN explosion of
a massive star (Gualandris et al.\ 2005). Recent observations with the
10-m Keck~II telescope revealed that the secondary star has a
supersolar surface metallicity ($[{\rm Fe/H}]=0.2\pm0.2$, Gonz\'alez
H\'ernandez et al. 2006), confirming the origin of the black hole
in a SN event. Thus, if the system originated in the Galactic
halo, the element abundances of the secondary star must have been
enriched by a factor of 5--25 depending on whether its initial 
metallicity resembled a thick-disk star or a halo star.

Element abundances of secondary stars of X-ray binaries have been
studied for the systems Nova Scorpii 1994 (Israelian et al. 1999;
Gonz\'alez Hern\'andez et al. 2007), A0620--00 (Gonz\'alez Hern\'andez
et al. 2004), Centaurus X-4 (Gonz\'alez Hern\'andez et al. 2005),
\mbox{XTE J1118+480} (Gonz\'alez Hern\'andez et al. 2006, hereafter
Paper~I), and V4641 Sagittarii (Orosz et al. 2001; Sadakane et
al. 2006). All of these X-ray binaries show metallicities close to
solar independent of their location with respect to the Galactic
plane, and possible scenarios of pollution from a SN or hypernova have
been discussed. In this paper, we compare in detail different
scenarios of the possible enrichment of the secondary star from SN
yields, providing conclusions on the formation region (Galactic halo,
thick disk, or thin disk) of this halo black hole X-ray binary.

\begin{figure}[ht!]
\centering
\includegraphics[width=6.2cm,angle=90]{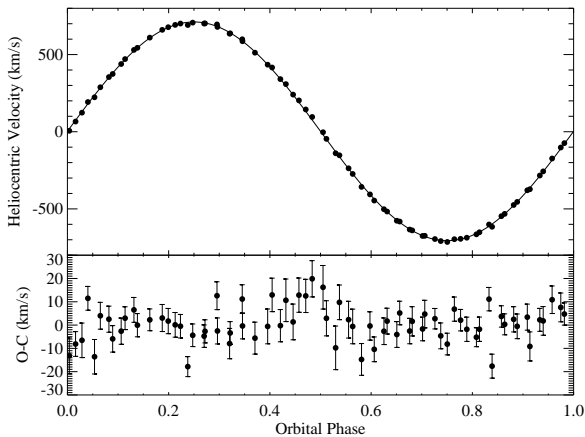}
\caption{\footnotesize{Radial velocities of \mbox{XTE J1118+480}
folded on the orbital solution of the data, together with the 
best-fitting sinusoid. Individual velocity uncertainties are $\le 7$ 
${\rm km}\ {\rm s}^{-1}$ and are not plotted because they are always 
smaller than the symbol size. The bottom panel shows the residuals
of the fit.}}   
\label{fig1}
\end{figure}

\section{Observations}

As already reported in Paper~I, we obtained 74 medium-resolution spectra 
of the black hole X-ray binary \mbox{XTE J1118+480}, in quiescence, with
the Echelle Spectrograph and Imager (ESI; Sheinis et al.\ 2002) at the
10-m Keck~II telescope on UT 14 February 2004. The data covered the
spectral range 4000--9000~{\AA} at a resolving power
$\lambda/\Delta\lambda \approx 6,000$. We also observed ten template 
stars with spectral types in the range K0V--M2V with the same
instrument and spectral configuration. The exposure time
was fixed at 300~s to minimize the effects of orbital smearing which, 
for the orbital parameters of \mbox{XTE J1118+480}, is in the
range 0.6--26.6 \kmso, smaller than the instrumental resolution of 50 
\kmso. All of the spectra were reduced in a standard manner.  

\begin{figure*}[ht!]
\centering
\includegraphics[width=11cm,angle=90]{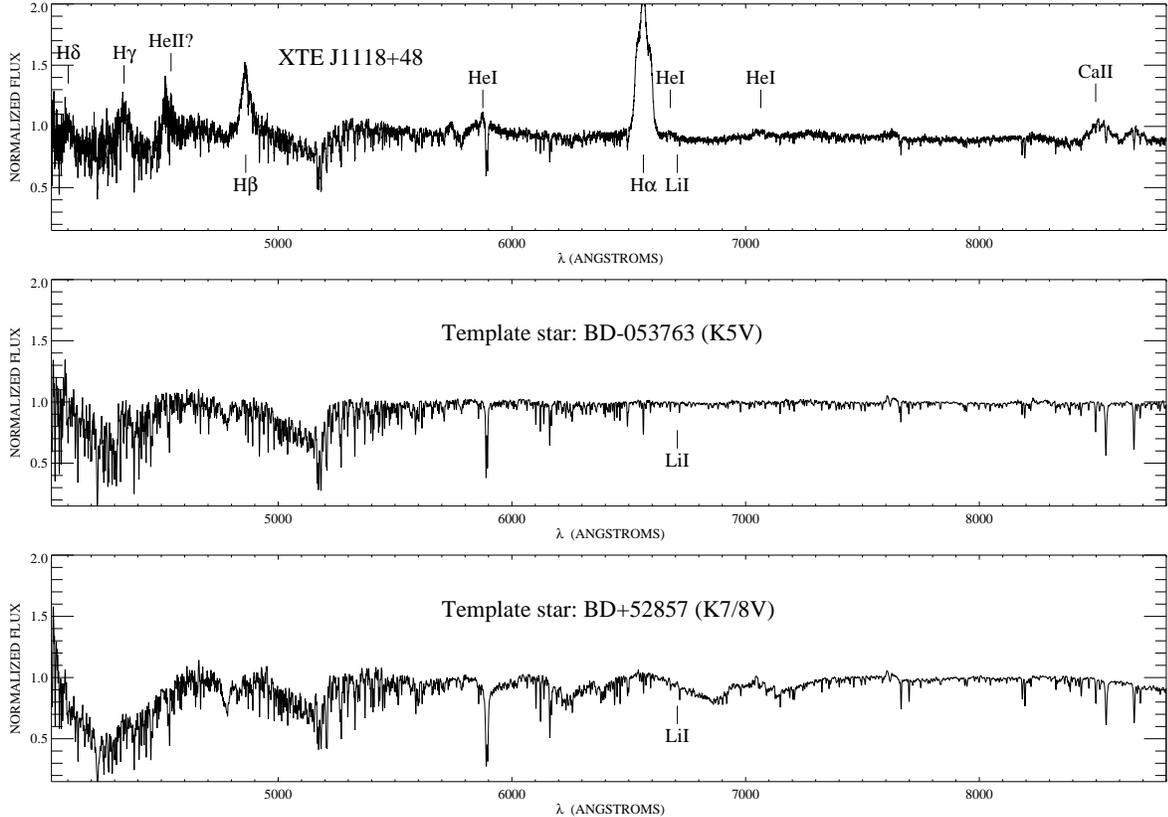}
\caption{\footnotesize{Observed spectrum of the secondary star of
\mbox{XTE J1118+480} (top panel) and of two properly broadened templates 
(BD$-$053763, middle panel; BD+52857, bottom panel).}}  
\label{fig2}
\end{figure*}

\section{Revised Orbital Parameters}

We extracted the radial velocities by cross-correlating each target
spectrum with the spectrum of a K5V template star, using the software
MOLLY developed by T. R. Marsh. We fitted these data with a sine wave
using a $\chi2$ method providing the following orbital solution (see
Fig.~\ref{fig1}): $\gamma= 2.7 \pm 1.1$ \kmso, $K_2 = 708.8 \pm 1.4$ \kmso,
$P=0.16995 \pm 0.00012$~d, and $T_0=2,453,049.93346 \pm 0.00007$~d,
where $T_0$ is defined as the corresponding time of the closest
inferior conjunction of the companion star, and the quoted
uncertainties are 1$\sigma$. This orbital period, $P$, and the 
velocity amplitude of the orbital motion of the secondary
star, $K_2$, lead to a mass function of $f(M) = 6.27 \pm 0.04$ \Msuno,
consistent with (but more precise than) previous results (McClintock 
et al. 2001; Wagner et al. 2001; Torres et al. 2004). 

The derived radial velocity of the center of mass of the
system agrees somewhat (at the 3$\sigma$ level) with previous
studies ($\gamma=+26\pm17$ \kmso, McClintock et al. 2001;
$\gamma=-15\pm10$ \kmso, Wagner et al. 2001; 
$\gamma=+16\pm6$ \kmso, Torres et al. 2004). Note that our
medium-resolution data have a factor of $\geq 4$ higher spectral
resolution than the data these authors used. We estimate the
uncertainty in our individual radial-velocity measurements of
typically 6~\kms (see Fig.~\ref{fig1}).  

\subsection{Secondary Spectrum}

The individual spectra were corrected for their radial velocity and 
combined in order to improve the signal-to-noise ratio (S/N). After
binning in wavelength in steps of 0.3~{\AA}, the final spectrum had 
S/N $\approx 80$ in the continuum in the red spectral
region. This spectrum, displayed in Fig.~\ref{fig2}, was compared with
ten template stars having spectral types K0V to M2V. The best fit shows
a K5V star rather than the later spectral types (K7/8V or even M)
suggested in previous studies (Wagner et al. 2001; Torres et al.
2004). 

We should remark that the spectrum of the secondary star in this
system and the spectra of the template stars were normalized using the
same procedure. The continuum was fitted with a low-order spline, 
in order to avoid the smoothing of possible existing broad TiO bands. 
Following Marsh et al. (1994), we computed the optimal value of
$v~\sin~i$ by subtracting broadened versions of the K5V template
(in steps of 1 \kmso) and minimizing the residual. We used a
spherical rotational profile with linearized limb-darkening $\epsilon
= 0.8$ (Al-Naimiy 1978) due to the spectral type K of the secondary
star. The best fit corresponds to $v~\sin~i=100^{+3}_{-11}$ \kmso,
where the uncertainties have been derived by assuming extreme cases for
$\epsilon = 0-1$. Our derived rotational velocity, combined with our
value of the velocity amplitude, $K_2$, implies a binary mass ratio
$q=0.027\pm0.009$, in agreement with previous results (Torres et al.
2004, and references therein).

\begin{deluxetable}{lccccc}
\tabletypesize{\scriptsize}
\tablecaption{Ranges and Steps of Model Parameters}   
\tablewidth{0pt}
\tablehead{\colhead{Parameter} & \colhead{Range} & \colhead{Step}}
\startdata
$T_{\mathrm{eff}}$  & $3500 \rightarrow 5000$ K & 100 K\\ 
$\log [g/({\rm cm~s}2)]$  & $4 \rightarrow 5$  & 0.1\\ 
$\mathrm{[Fe/H]}$ & $-1.5 \rightarrow 1$  & 0.1\\ 
$f_{4500}$ &  $0 \rightarrow 2$  & 0.1\\ 
$m_0$  & $0 \rightarrow -0.00091$ & $-$0.00010\\ 
\enddata
\label{tbl1}
\end{deluxetable}


\begin{figure*}[ht!]
\centering
\includegraphics[width=11cm,angle=0]{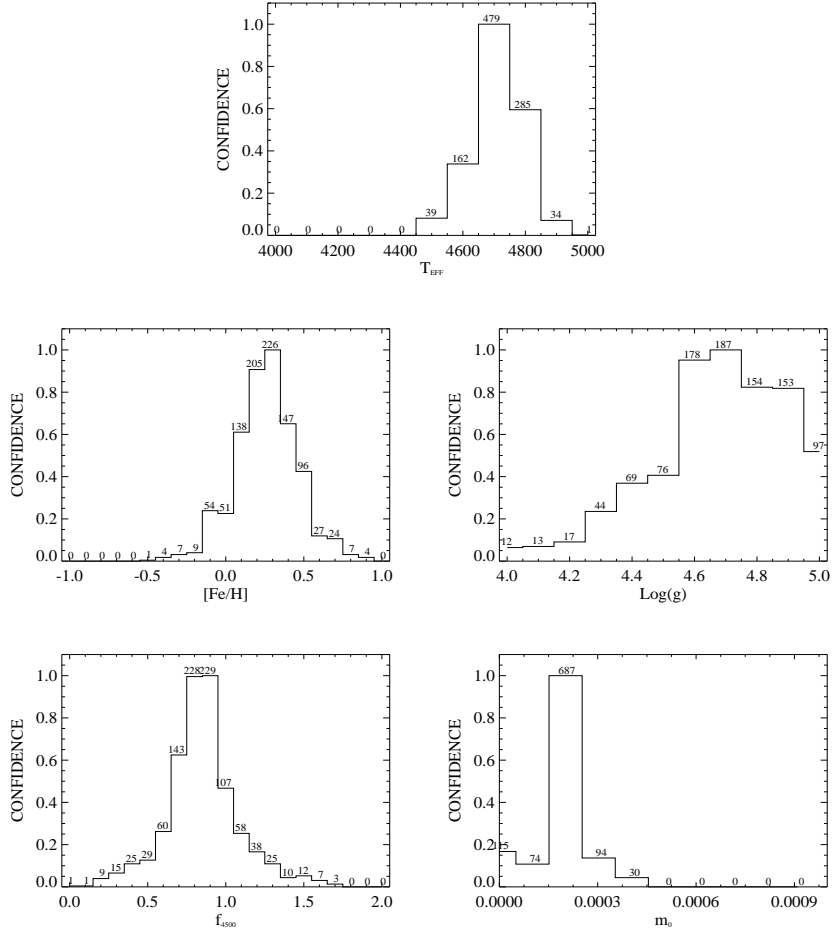}
\caption{\footnotesize{Distributions obtained for each parameter using
Monte Carlo simulations. The labels at the top of each bin indicate
the number of simulations consistent with the bin value. The total
number of simulations was 1000.}}
\label{fig3}
\end{figure*}

\begin{figure*}[ht!]
\centering
\includegraphics[width=9cm,angle=90]{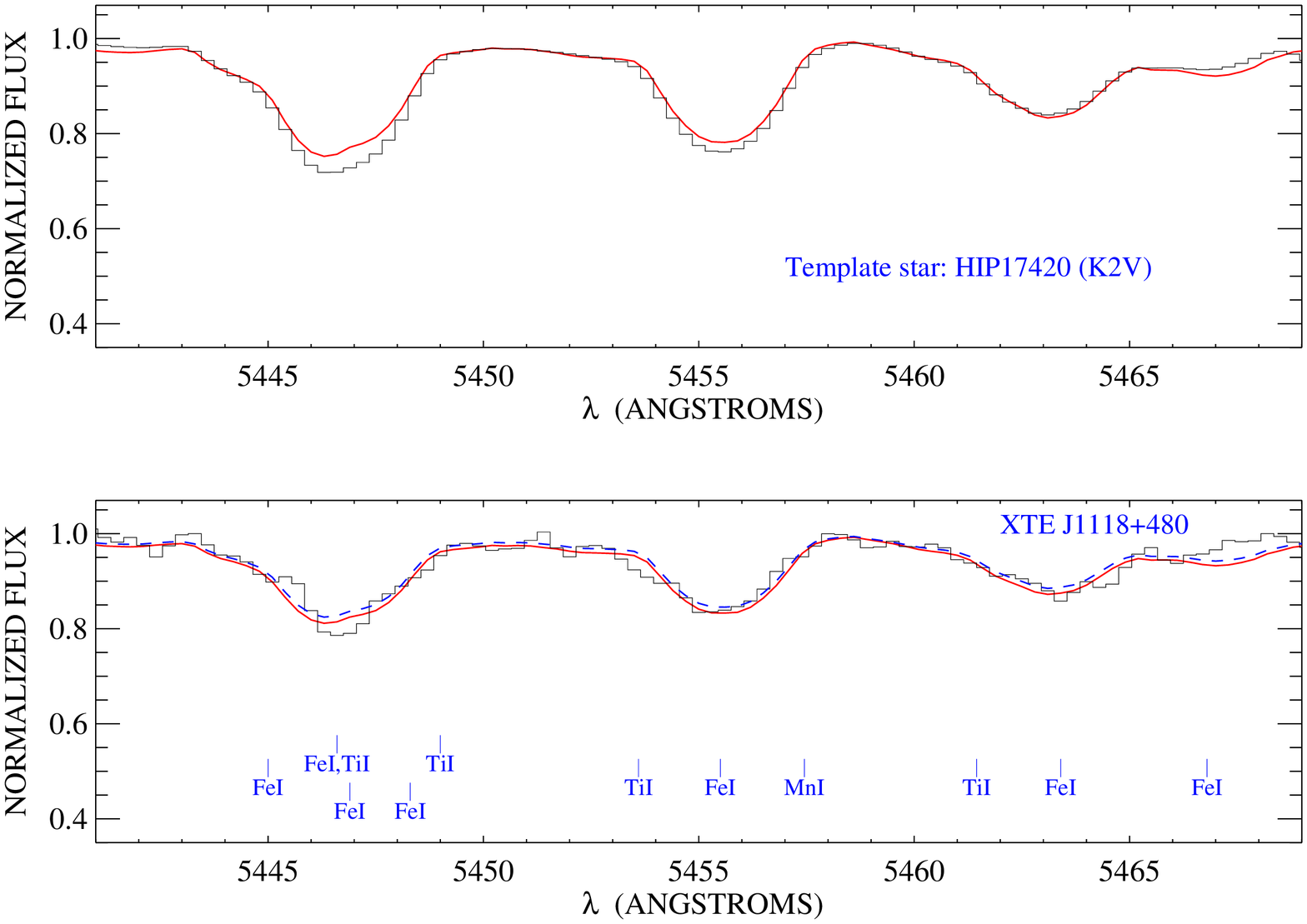}
\caption{\footnotesize{Best synthetic spectral fits to the Keck/ESI
spectrum of the 
secondary star in the \mbox{XTE J1118+480} system (bottom panel) and
the same for a template star (properly broadened) shown for comparison
(top panel). Synthetic spectra are computed for solar abundances
(dashed line) and best-fit abundances (solid line).}}
\label{fig4}
\end{figure*}

\begin{figure*}[ht!]
\centering
\includegraphics[width=9cm,angle=90]{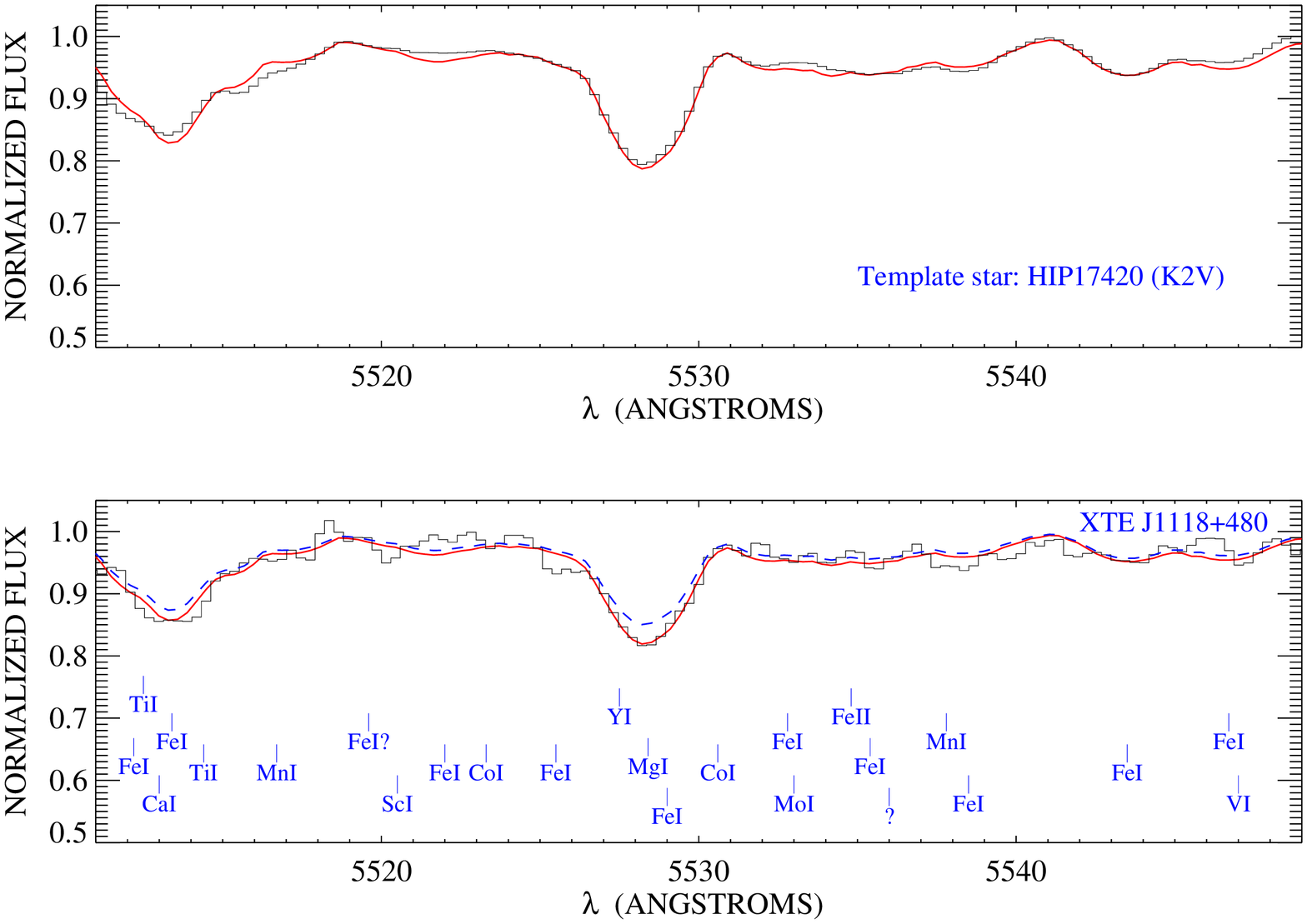}
\caption{\footnotesize{The same as in Fig.\ 4, but for the spectral range
5510--5550 {\AA}.}}  
\label{fig5}
\end{figure*}

\begin{figure*}[ht!]
\centering
\includegraphics[width=9cm,angle=90]{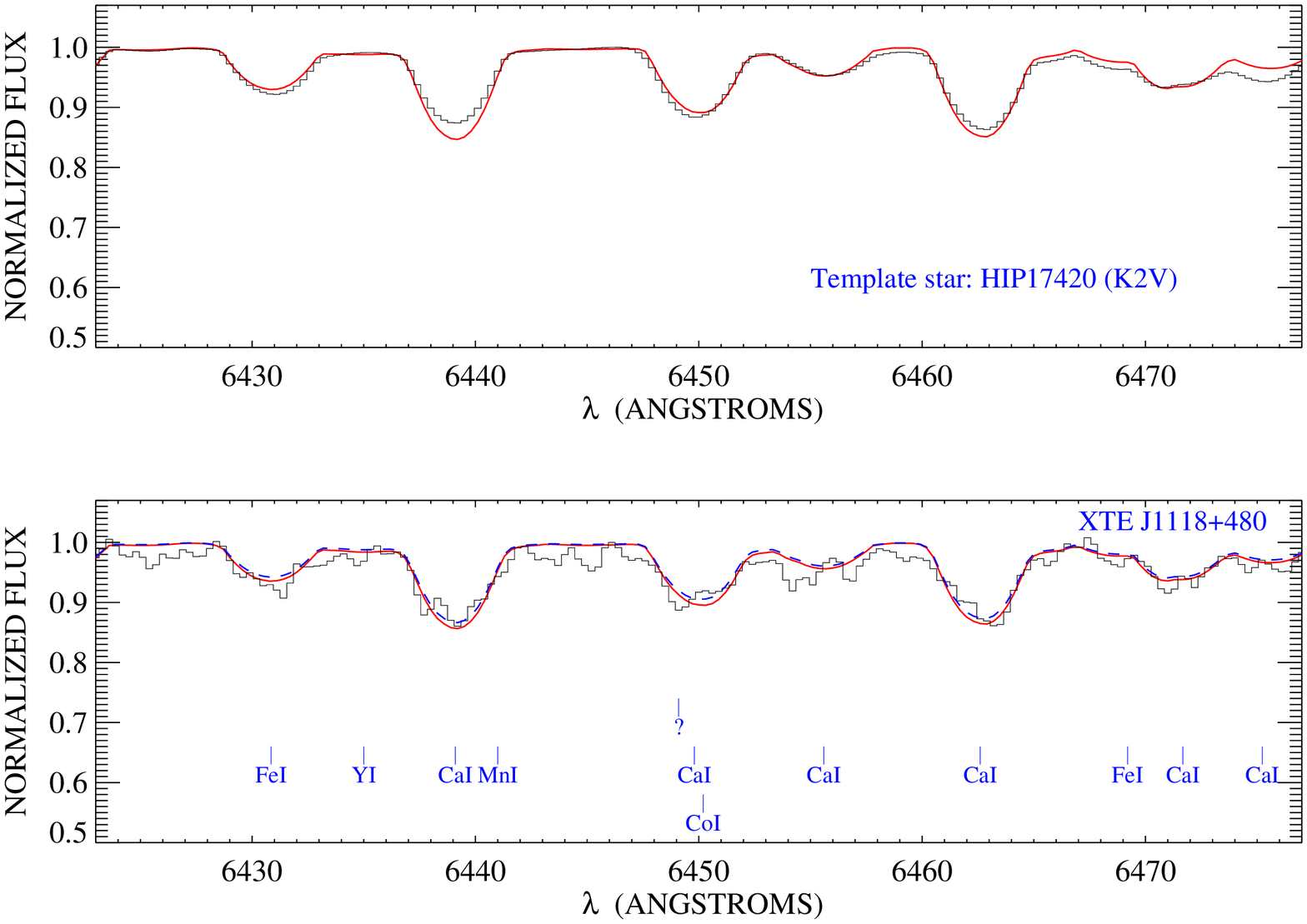}
\caption{\footnotesize{The same as in Fig.\ 4, but for the spectral range
6420--6480 {\AA}.}}  
\label{fig6}
\end{figure*}

\section{Chemical Analysis}

\subsection{Stellar Parameters}

The normalized spectra of X-ray transients, although observed in
quiescence, show apparently weaker stellar lines of the secondary star
due to the veiling caused by the accretion disk. The veiling from the
accretion disk was estimated to be $\sim65\% \pm 8\%$ in the spectral 
range 5800--6400~{\AA} in December 2000 and January 2001, and
$\sim 40\% \pm 10\%$ in January 2003, by performing standard optimal
subtraction techniques with K5V--M0V template stars. 

As shown in Paper~I, 
we tried to infer the stellar parameters, $T_{\mathrm{eff}}$ and $\log
g$, and the metallicity [Fe/H], of the secondary star taking into account
the veiling from the accretion disk, defined as a linear function of
wavelength and thus described with two additional parameters, the
veiling at 4500~{\AA}, $f_{4500} = F^{4500}_{\rm disk}/F^{4500}_{\rm
sec}$, and the slope, $m_0$. Note that the total flux is defined as
$F_{\rm total} = F_{\rm disk} + F_{\rm  sec}$, where $F_{\rm disk}$ and
$F_{\rm sec}$ are the flux contributions of the disk and the continuum
of the secondary star, respectively. This procedure involves a
$\chi2$ minimization routine which compares several features of the 
stellar spectrum with a grid of synthetic spectra computed using the
LTE code MOOG (Sneden 1973). We used a grid of LTE model atmospheres
(Kurucz 1993) and atomic line data from the Vienna Atomic Line
Database (VALD, Piskunov 1995). The oscillator strengths of relevant
lines were adjusted until they reproduced the solar atlas (Kurucz et al.\
1984) with solar abundances (Grevesse et al. 1996). The changes we
applied to the $\log gf$ values taken from the VALD database were
$\Delta\log gf \lesssim 0.2$ dex. 

\begin{deluxetable}{lccccccc}
\tabletypesize{\scriptsize}
\tablecaption{Abundance Uncertainties in \mbox{XTE J1118+480}}
\tablewidth{0pt}
\tablehead{\colhead{Element} & $\mathrm{[X/H]}_{\rm
LTE}$\tablenotemark{a} & $\Delta_{\sigma}$ &
$\Delta_{T_{\mathrm{eff}}}$ & $\Delta_{\log g}$
& $\Delta_{\rm veil}$ & $\Delta_{\rm
tot}$\tablenotemark{b} & $n_{\rm lines}$\tablenotemark{c}}
\startdata
Mg   &  0.35 & 0.12 & 0    & $-$0.10  & 0.20 & 0.25 & 1\\
Al   &  0.60 & 0.12 & 0.05 &  0     & 0.15 & 0.20 & 1\\
Si   &  0.37 & 0.03 & $-$0.07&  0.07  & 0.18 & 0.21 & 2\\
Ca   &  0.15 & 0.03 & 0.13 & $-$0.16  & 0.11 & 0.23 & 5 \\
Ti   &  0.32 & 0.18 & 0.12 & $-$0.03  & 0.14 & 0.26 & 3\\
Fe   &  0.18 & 0.08 & 0.06 &  0.04  & 0.13 & 0.17 & 5 \\
Ni   &  0.30 & 0    & 0.10 &  0.12  & 0.14 & 0.21 & 2 \\
Li\tablenotemark{d} & 1.78 & 0.12 & 0.15 & 0.05 & 0.15  & 0.25  & 1\\
\enddata

\tablenotetext{a}{Element abundances of the secondary star (calculated 
assuming LTE) are $\mathrm{[X/H]}= \log [N(\mathrm{X})/N(\mathrm{H})]_{\rm
star} - \log [N(\mathrm{X})/N(\mathrm{H})]_{\rm Sun}$, where
$N(\mathrm{X})$ is the number density of atoms. Uncertainties, $\Delta
\mathrm{[X/H]}$, are at the 1$\sigma$ level and take into account the
uncertainties in the stellar and veiling parameters.}

\tablenotetext{b}{The total error was estimated as $\Delta_{\rm tot} 
= \sqrt{\Delta^2_{\sigma} + \Delta^2_{T_{\mathrm{eff}}} + 
\Delta^2_{\log g} + \Delta^2_{\rm veil}}.$} 

\tablenotetext{c}{Number of features analyzed for each
element.} 

\tablenotetext{d}{\mbox{Li} abundance is expressed as $\log 
\epsilon(\mathrm{Li})_{\rm NLTE} = \log 
[N(\mathrm{Li})/N(\mathrm{H})]_{\rm NLTE} + 12$.}

\tablecomments{The uncertainties from the dispersion of the best
fits to different features, $\Delta_{\sigma}$, are estimated using the
following formula: $\Delta_{\sigma} =\sigma/\sqrt{N}$, where $\sigma$
is the standard deviation of the measurements.}

\label{tbl2}
\end{deluxetable}

We selected nine spectral features containing in total 30 lines of
Fe~I and 8 lines of Ca~I with excitation potentials between 1 and 5~eV. 
The five free parameters were varied in the ranges given in Table~\ref{tbl1}.
For each given iron abundance in the range $[{\rm Fe}/{\rm H}] < 0$, the
Ca abundance was fixed according to the Galactic trend of Ca (Bensby
et al.\ 2005), while for $[{\rm Fe}/{\rm H}] > 0$, we assumed $[{\rm
Ca}/{\rm Fe}]=0$. A rotational broadening of 100 {\kms} and a
limb darkening $\epsilon = 0.8$ were adopted. The microturbulence,
$\xi$, was computed using an experimental expression as a function of
effective temperature and surface gravity (Allende Prieto et al.\
2004).   
 
The result, already presented in Paper~I, provides as most likely
values $T_{\mathrm{eff}} = 4700 \pm 100$~K, $\log [g/{\rm cm~s}^2] =
4.6 \pm 0.3$, $\mathrm{[Fe/H]} = 0.18 \pm 0.17$, $f_{4500} =
0.85 \pm 0.20$, and $m_0 = -0.0002 \pm 0.0001$. The 1$\sigma$
uncertainties of the five free parameters were determined using 1000
realizations whose corresponding histograms are displayed in
Fig.~\ref{fig3}. 

\begin{figure*}[ht!]
\centering
\includegraphics[width=13.5cm,angle=0]{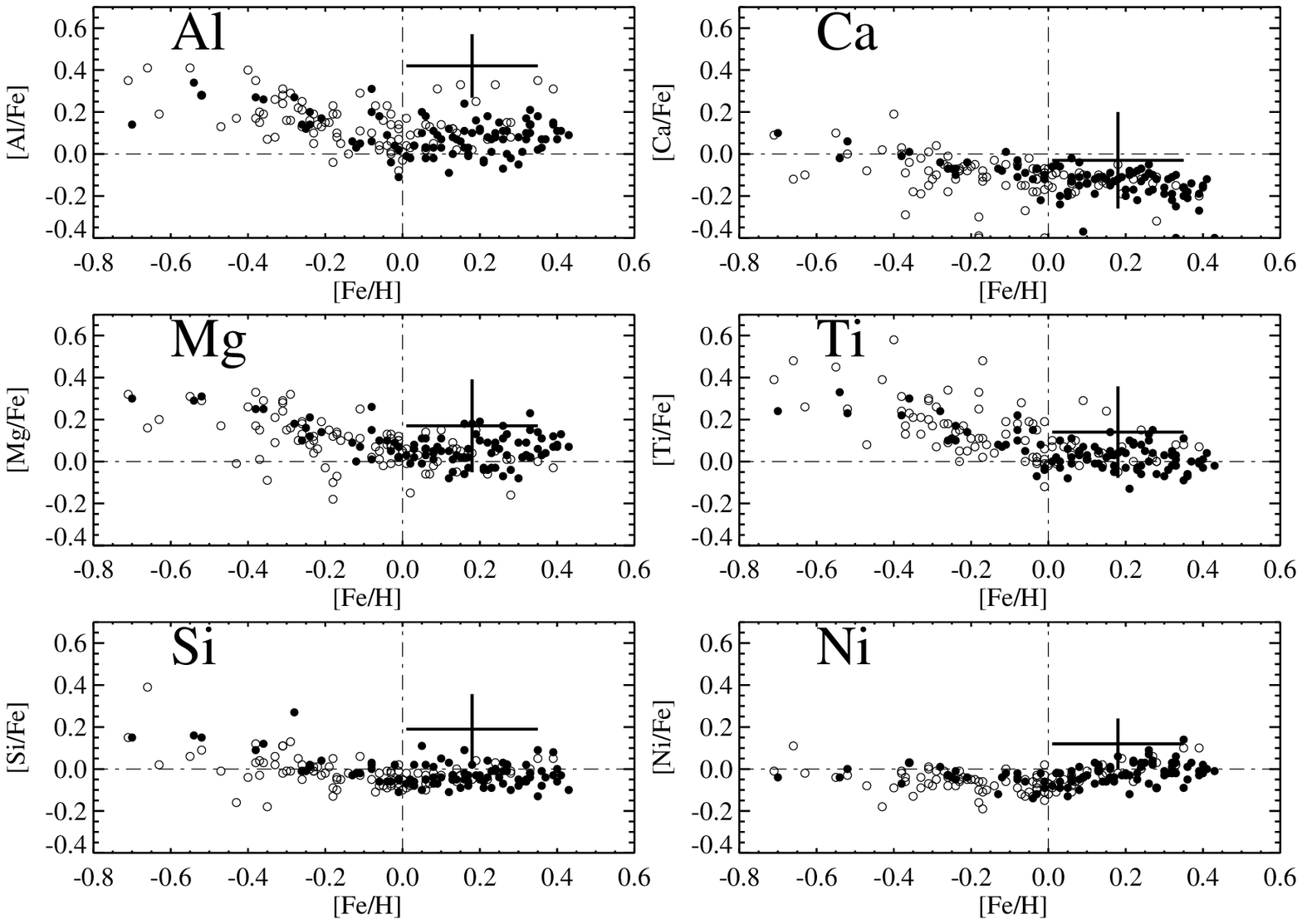}
\caption{\footnotesize{Abundance ratios of the secondary star in
\mbox{XTE J1118+480} (blue wide cross) in comparison with the
abundances of G and K metal-rich dwarf stars. Galactic trends were
taken from Gilli et al. (2006). The size of the cross 
indicates the uncertainty. Filled and empty circles correspond to
abundances for exoplanet host stars and stars without known exoplanet
companions, respectively. The dashed-dotted lines indicate solar
abundance values.}} 
\label{fig7}
\end{figure*}

The stellar parameters derived, especially the effective temperature,
could provide important implications on the Gelino et al. (2006)
determination of the orbital inclination of the system ($i \approx 
68^\circ$). These authors modeled the optical and infrared (IR)
ellipsoidal light curves of \mbox{XTE J1118+480} in quiescence,
assuming a K7V spectral type (\teff $\approx 4250$~K) for the secondary
star. However, our spectroscopic value (\teff $\approx 4700$~K) would
require more contribution of the flux from the accretion disk in the $K$
band. Gelino et al. (2001) derived an inclination of $i \approx 41^\circ$,
yielding a black hole mass of $M_{\rm BH} \approx 11$~\Msun for 
the \mbox{A0620--00} system, by adopting \teff $\approx 4600$~K, 
which is 300~K lower than the spectroscopic value
reported by Gonz\'alez Hern\'andez et al. (2004). Hynes et al. (2005)
suggested that this different effective temperature of the 
secondary star in \mbox{A0620--00} would require a larger disk
contribution in the $K$ band, and therefore a higher inclination. In
fact, Gelino et al. (2001) commented that if the $K$-band disk veiling
as high as 50\% of the total flux, the derived inclination would increase to
60$^\circ$. The black hole mass would then drop to $M_{\rm BH} \approx 
5$ \Msuno. Although this is an extreme case, milder IR veiling could
still have a substantial impact on the derived black hole mass.
Similarly, the black hole mass in \mbox{XTE J1118+480} might be 
significantly affected when using our value of the effective temperature.

\subsection{Stellar Abundances}
\begin{deluxetable}{lccccc}
\tabletypesize{\scriptsize}
\tablecaption{Element Abundance Ratios in \mbox{XTE J1118+480}}
\tablewidth{0pt}
\tablehead{\colhead{Element} & $\mathrm{[X/Fe]}_{\rm XTE1118}$ &
$\Delta^{\tablenotemark{\star}}_{\rm [X/Fe],XTE1118}$ & $\mathrm{[X/Fe]}_{\rm stars}$ & $\sigma_{\rm
stars}$ & $\Delta_{\sigma,{\rm stars}}$} 
\startdata
Mg   &  0.17 & 0.22 &  0.02 & 0.08 & 0.02 \\
Al   &  0.42 & 0.15 &  0.14 & 0.10 & 0.02 \\
Si   &  0.19 & 0.17 & $-$0.03 & 0.04 & 0.01 \\
Ca   & $-$0.03 & 0.23 & $-$0.13 & 0.05 & 0.01 \\
Ti   &  0.14 & 0.22 &  0.05 & 0.08 & 0.01 \\
Ni   &  0.12 & 0.12 & $-$0.02 & 0.05 & 0.01 \\
\enddata

\tablenotetext{\star}{Uncertainties in the element abundance ratios
($\mathrm{[X/Fe]}$) in the secondary star in \mbox{XTE J1118+480}.}

\tablecomments{$\mathrm{[X/Fe]}_{\rm stars}$ indicate the average
values of 24 stars with iron content in the range 0.01 to 0.35
corresponding to 1$\sigma$ in the $\mathrm{[Fe/H]}$ abundance of the
secondary star in \mbox{XTE J1118+480}, taken from Gilli et al.
(2006). The uncertainty in the average value of abundance
ratios in the comparison sample is obtained as $\Delta_{\sigma,{\rm
stars}} =\sigma_{\rm stars}/\sqrt{N}$, where $\sigma_{\rm stars}$ is
the standard deviation of the measurements and N is the number of stars.}    

\label{tbl3}
\end{deluxetable}

We inspected several spectral regions in the observed Keck/ESI spectrum 
of the secondary star, searching for suitable lines for a detailed
chemical analysis. Using the derived stellar parameters, we first 
determined the Fe abundance by comparing synthetic spectra with each
individual feature in the ESI spectrum (see Table~\ref{tbl2}). In
Fig.~\ref{fig4} (here) and Fig.~1 of Paper~I, we display some of the spectral
regions analyzed to obtain the Fe abundance. This figure also shows the
best synthetic spectral fit to the observed spectrum of a template
star (HIP 17420 with \teffo$=4801$ K, \loggo$ = 4.633$, and
[Fe/H]$=-0.14$ dex) using the stellar parameters and abundances
determined by Allende-Prieto et al. (2004). We only use as abundance
indicators those features which were well reproduced in the template
star. The chemical analysis is summarized in Table~\ref{tbl2}. The
errors in the element abundances show their sensitivity to the
uncertainties in the effective temperature
($\Delta_{T_{\mathrm{eff}}}$), gravity ($\Delta{\log g}$), veiling
($\Delta_{\rm veiling}$), and the dispersion of the measurements
from different spectral features ($\Delta_{\sigma}$). In
Table~\ref{tbl2} we also state the number of features analyzed for
each element.  

\begin{deluxetable}{lcccccccc}
\tabletypesize{\scriptsize}
\tablecaption{Supernova/Hypernova Model Parameters}
\tablewidth{0pt}
\tablehead{ $M_{\rm MS}$ & $M_{\rm He}$ & $E_{51}$ & 
\colhead{Z} & $M(Fe)$ & $M_{\rm cut}$ &
$M_{\rm fall}$ & 
$M_{{\rm BH},i}$\tablenotemark{a} & \colhead{Abundance Pattern}}
\startdata
40   & 15.8 & 1  & 0.001 & 5.8982E-01 & 2.51 & 5.56 & 8.07 & Fig.~9a \\
40   & 15.8 & 1  & 0.001 & 2.6581E-01 & 3.01 & 5.06 & 8.07 & Fig.~9a \\
40   & 15.8 & 30 & 0.001 & 7.6394E-01 & 4.05 & 4.02 & 8.07 & Fig.~9b \\
40   & 15.8 & 30 & 0.001 & 9.2867E-02 & 5.03 & 3.04 & 8.07 & Fig.~9b \\
40   & 15.8 & 1  & 0.004 & 3.7283E-01 & 2.50 & 5.57 & 8.07 & Fig.~10a \\
40   & 15.8 & 1  & 0.004 & 2.6649E-02 & 3.03 & 5.04 & 8.07 & Fig.~10a \\
40   & 15.8 & 30 & 0.004 & 9.8590E-01 & 3.03 & 5.04 & 8.07 & Fig.~10b \\
40   & 15.8 & 30 & 0.004 & 2.0354E-01 & 4.02 & 4.05 & 8.07 & Fig.~10b \\
30   & 11   & 1  & 0.004 & 1.4667E-01 & 2.52 & 5.49 & 8.01 & Fig.~11a \\
30   & 11   & 1  & 0.004 & 1.7806E-03 & 3.04 & 4.97 & 8.01 & Fig.~11a \\
30   & 11   & 20 & 0.004 & 5.7203E-01 & 3.04 & 4.97 & 8.01 & Fig.~11b \\
30   & 11   & 20 & 0.004 & 3.0448E-02 & 4.15 & 3.86 & 8.01 & Fig.~11b \\
40   & 15.1 & 1  & 0.02 & 7.3775E-01 & 1.46 & 6.63 & 8.09 & Fig.~12a \\
40   & 15.1 & 1  & 0.02 & 3.1602E-03 & 8.09 & 0    & 8.09 & Fig.~12a \\
40   & 15.1 & 30 & 0.02 & 8.4572E-01 & 1.74 & 6.35 & 8.09 & Fig.~12b \\
40   & 15.1 & 30 & 0.02 & 2.8515E-03 & 8.09 & 0    & 8.09 & Fig.~12b \\
40   & 16 & 10 & 0.02 & 1.4337E-02 & 2.41 & 5.15 & 7.56 & Fig.~13a \\
40   & 16 & 10 & 0.02 & 0	    & 7.56 & 0    & 7.56 & Fig.~13a \\
40   & 16 & 10 & 0.02 & 5.1974E-01 & 2.41 & 5.15 & 7.56 & Fig.~13b \\
40   & 16 & 10 & 0.02 & 1.2716E-01 & 7.56 & 0    & 7.56 & Fig.~13b \\
\enddata

\tablenotetext{a}{$M_{{\rm BH},i}$ is the mass of the remnant after the
explosion and before the secondary started to transfer matter onto the
compact object. $M_{{\rm BH},f}=8.53\pm0.60$ \Msun is the observed
mass of the black hole.}

\tablecomments{Supernova and hypernova explosion models used in
Figs.~9-13. The quantities shown are the main-sequence mass, the mass
of the He core, the explosion energy, $E_{51}=E_K/10^{51} \,{\rm
erg}$, the ejected Fe mass, the metallicity of the model, the
mass-cut, the mass of the fallback matter, and the final remnant mass.
The mass are in units of \Msuno.}

\label{tbl4}
\end{deluxetable}

The abundances of Ti and Si were mainly derived from several
lines in the spectral region 5920--5960 {\AA} where some telluric lines
are present. However, since 74 spectra with different radial
velocities in the range $\pm710$ \kms were combined to generate the
average spectrum of the secondary star, these telluric lines must have
been smoothed out. 

The Mg abundance was derived from one spectral line (see
Fig.~\ref{fig5}) and the error associated with the dispersion of the
measurements, $\sigma$, was assumed to be the average dispersion of  
Fe, Ca, and Ni abundances, and in this case, $\Delta_{\sigma}=\sigma$.
The same prescription was adopted for the analysis of Al and Li (see
Fig.~1 of Paper~I). The best fit to the \ion{Li}{1} 6708~{\AA} feature
provides an LTE abundance of $\log \epsilon(\mathrm{Li})_{\rm
LTE}=1.61 \pm 0.25$. We estimated the non-LTE abundance
correction\footnote{$\Delta_\mathrm{NLTE}=\log
\epsilon(\mathrm{X})_\mathrm{NLTE}-\log
\epsilon(\mathrm{X})_\mathrm{LTE}.$} for this element from the 
theoretical LTE and non-LTE curves of growth in Pavlenko \& Magazz\`u
(1996). We found $\Delta_\mathrm{NLTE}= 0.17$. Due to the weakness of
the absorption we consider this abundance estimate given in
Table~\ref{tbl2} as an upper limit. 

\begin{figure*}[ht!]
\centering
\includegraphics[width=16.cm,angle=0]{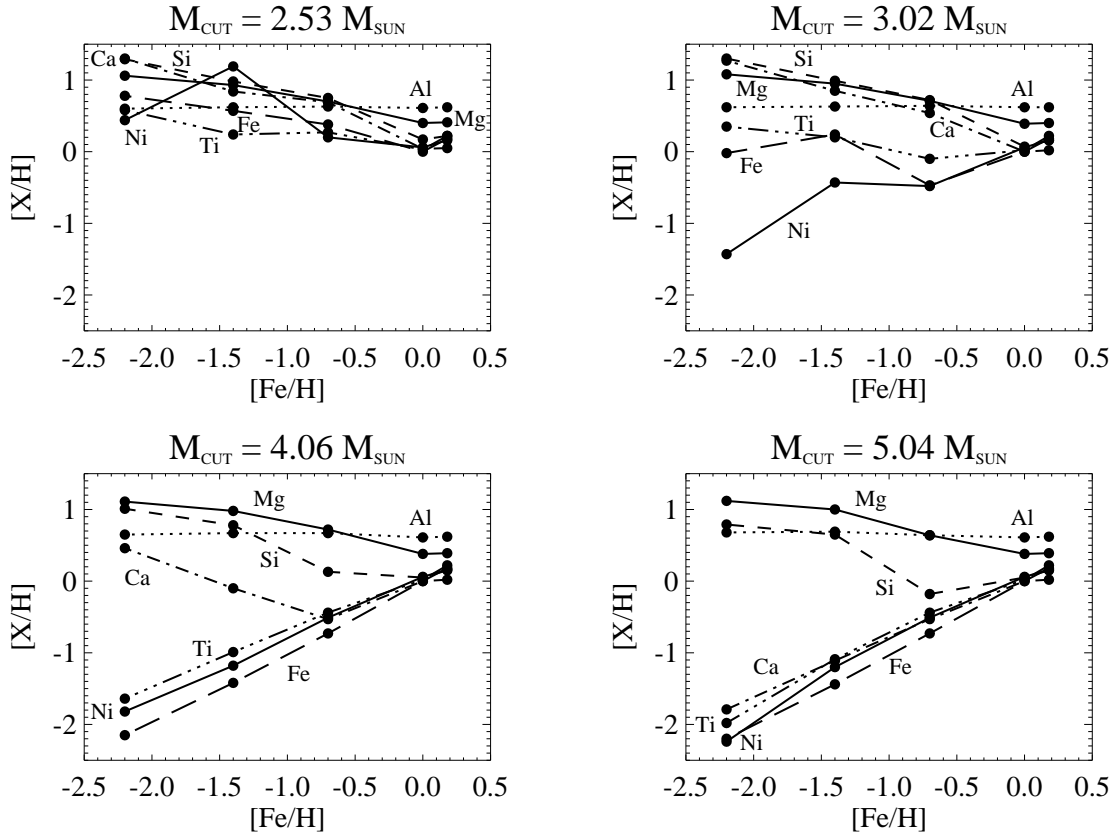}
\caption{\footnotesize{Chemical composition of the secondary
atmosphere contaminated with nucleosynthetic products of a 40~\Msun
spherically symmetric core-collapse SN explosion model ($M_{\rm He}
\approx 15$--16~\Msuno) for different metallicities and mass-cuts, $M_{\rm
cut}$.}} 
\label{fig8}
\end{figure*}

\section{Discussion}

As already discussed in Paper~I, the Fe abundance of the secondary
star is slightly higher than solar, but similar to that of many stars
in the solar neighborhood. The abundances of other elements listed in
Table~\ref{tbl2} relative to iron are compared in Fig.~\ref{fig7} with
the Galactic trends of these elements in the relevant range of
metallicities. Moderate anomalies are found only for Al. In
Table~\ref{tbl3} we show the element abundance ratios in the secondary
star in \mbox{XTE J1118+480} and the average values in stars with
iron content in the range $0.01 < \mathrm{[Fe/H]} < 0.35$, the
comparison sample, corresponding to a 1$\sigma$ uncertainty in the
iron abundance of the companion star. Whereas Ca and Ti are consistent
with the average values of the comparison sample, Ni and Si, at
1$\sigma$, and especially Al, at 2$\sigma$, appear to be more abundant
than the average values of the stars in the comparison sample.        

The present location and space-velocity components ($U$, $V$) of the
system might suggest that the system belongs to the Galactic halo, but
the derived metallicity makes this possibility less likely. One could 
include the metallicity in the expression given in Bensby et al. (2005) to
estimate the relative likelihoods that a star belongs to the Galactic
thin disk, thick disk, and halo. The equations could be written as
follows:

\small
\begin{eqnarray}
P_{\rm thin-disk}=f_{\rm D}\frac{P_{\rm D}}{P},\nonumber\\
P_{\rm thick-disk}=f_{\rm TD}\frac{P_{\rm TD}}{P},\nonumber\\
P_{\rm halo}=f_{\rm H}\frac{P_{\rm H}}{P},\nonumber\\
P_i=K_i \times \exp\left[-\frac{U2_{\rm
LSR}}{2\sigma2_{U_i}}
-\frac{(V_{\rm LSR}-V_{{\rm asym},i})2}{2\sigma2_{V_i}}\right]\nonumber\\
 \times \exp\left[-\frac{W2_{\rm LSR}}{2\sigma2_{W_i}}
-\frac{({\rm [Fe/H]}-{\rm [Fe/H]}_{{\rm asym},i})2}
{2\sigma2_{\rm [Fe/H]_i}}\right],\nonumber\\
P=f_{\rm D}P_{\rm D}+f_{\rm TD}P_{\rm TD}+f_{\rm H}P_{\rm H},\nonumber\\
~~{\rm and}~~ K_i=\frac{1}{(2\pi)2\sigma_{U_i}\sigma_{V_i}\sigma_{W_i}\sigma_{{\rm [Fe/H]}_i}},\nonumber  
\end{eqnarray}
\normalsize

\noindent
where the subscript $i$ indicates the three populations $D$
(thin disk), $TD$ (thick disk), and $H$ (halo). The total probability,
$P$, takes into account the fraction of stars belonging to each
population in the solar neighborhood ($f_{\rm D}=0.94$, $f_{\rm TD}=0.06$, 
and $f_{\rm H}=0.0015$; Bensby et al. 2003). The velocity distributions of each
population with respect to the local standard of rest (LSR) are
centered at zero except for the component $V_{\rm LSR}$, whose
center is displaced according to $V_{{\rm asym},i}$ (with $V_{{\rm
asym},D}=-15$\kmso, $V_{{\rm asym},TD}=-46$ \kms, and $V_{{\rm
asym},H}=-220$ \kmso). The metallicity distributions have been
characterized with ${\rm [Fe/H]}_{\rm asym,D}=-0.1$, ${\rm
[Fe/H]}_{\rm asym,TD}=-0.7$, and ${\rm [Fe/H]}_{\rm asym,H}=-1.4$, 
as well as with $\sigma_{{\rm [Fe/H]}_{\rm D}}=0.2$, 
$\sigma_{{\rm [Fe/H]}_{\rm TD}}=0.24$, and
$\sigma_{{\rm [Fe/H]}_{\rm H}}=0.5$, according to Allende Prieto et al.
(2004, 2006). Thus, the probability that a star with the Galactic
space velocity components of this system (Mirabel et al.\ 2001;
$U=-105\pm16$ {\kms}, $V=-98\pm16$ {\kms}, $W=-21\pm10$ {\kms}) 
and metallicity $\mathrm{[Fe/H]} = 0.18$ belongs to the Galactic halo 
is less than 0.1\%. Moreover, the kinematics alone suggest thick-disk
rather than halo membership, although the high metallicity of the
secondary star favors thin-disk membership. However, the system could
also have originated in a satellite galaxy. In particular, its
galactocentric orbit (Mirabel et al. 2001) is marginally consistent
with the equatorial orbit of the stream of the dwarf galaxy in Canis
Major, although its present Galactic latitude differs by more than
$40^\circ$ from the $l-b$ distribution of the remnant of this dwarf
galaxy, accreted by the Milky Way (Martin et al. 2004). 

 The system could also have originated as a consequence of an
encounter of an ancient black hole of the Galactic halo with a 
single star or a binary of two solar-type stars of the Galactic disk. 
However, this possibility is very unlikely due to the extremely low 
density of stars in the disk ($\sim$0.006 stars/pc$^3$; Mihalas
\& Binney 1981). The orbit of the system integrated backward in time
never crossed the Galactic plane through the inner 2~kpc (Gualandris
et al. 2005), so high-density regions near the Galactic center are
discarded. Portegies Zwart et al. (1997b) have modeled the encounter
of black holes in high-density systems. In high-density systems, where
the density of stars is $\sim4 \times 10^6$ stars/pc$^3$ ($10^9$ times higher 
than in the Galactic disk), a black hole spends 1.5~Gyr before it suffers 
a tidal capture by a main-sequence star. In addition, the cases where a
black hole can capture an isolated star or one star of a binary require
very stringent constraints on the closest approach and impact velocity
(Benz \& Hills 1992; Hills 1991). All these reasons make this
possibility very unlikely.

Therefore, in conclusion, the present location, velocity, and
metallicity of the secondary star in \mbox{XTE J1118+48} suggest that
the black hole formed in a supernova/hypernova explosion that occurred
within the binary system. This explosive event must have either
provided a kick to the system if it was formed in the thin disk, or
enriched significantly the atmosphere of the secondary star if the
system formed in the thick disk or halo. 

The present orbital separation between the compact object and the
secondary star has been estimated to be $a_{\rm c} \approx
2.67$ {$R_\odot$} (Gelino et al. 2006). Thus, the secondary star could
have captured a significant amount of the ejected matter in the
SN explosion that formed the compact object. The chemical
composition of the secondary star may provide information on the
chemical composition of the progenitor of the compact object, and
therefore on the formation region (thin disk, thick disk, or halo) of
the binary system. We will now discuss the possibility that the SN
explosion of the massive progenitor enriched the secondary star from
different initial metallicities.

\begin{figure*}[ht!]
\centering
\includegraphics[width=8.cm,angle=0]{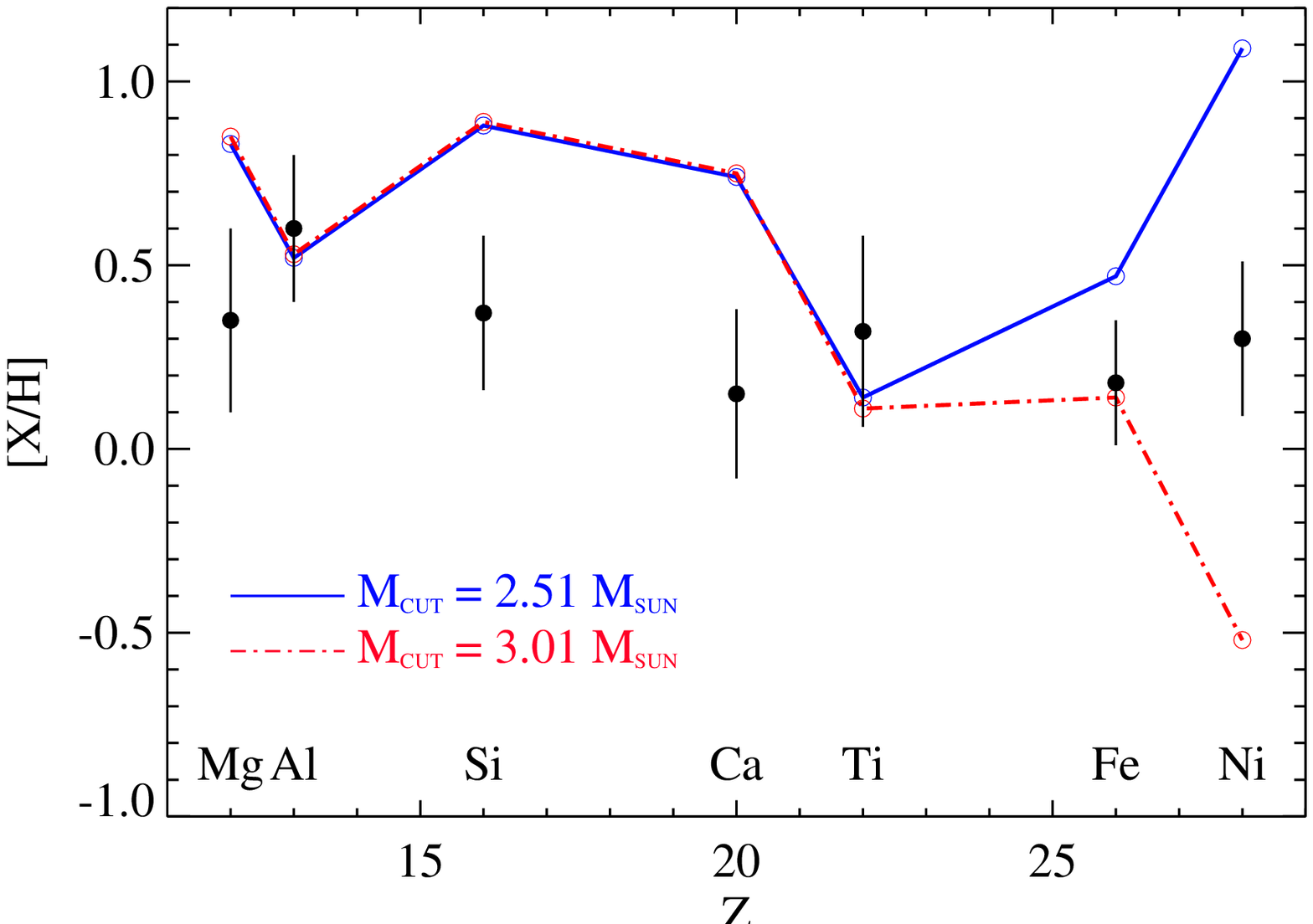}
\includegraphics[width=8.cm,angle=0]{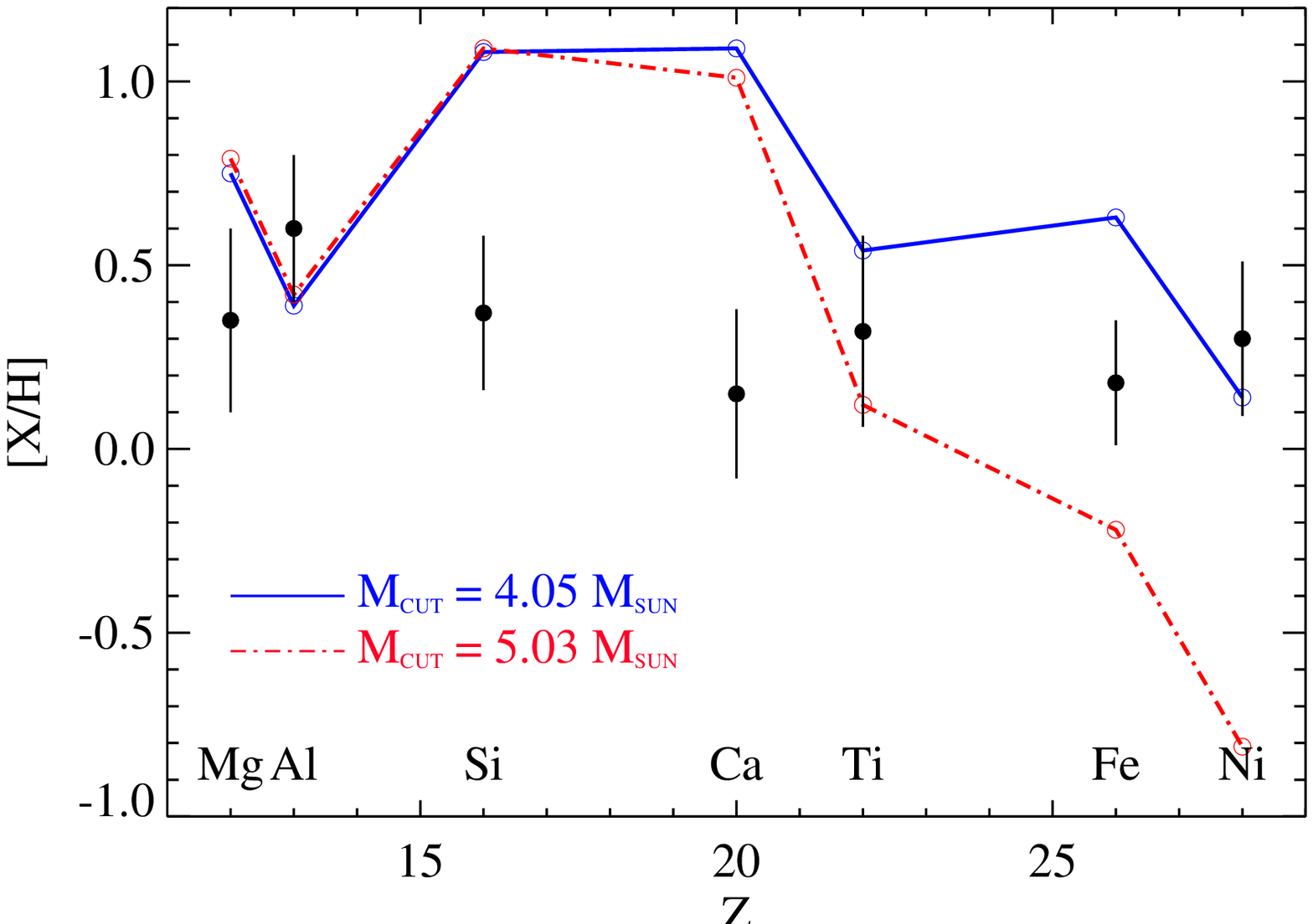}
\caption{\footnotesize{Left panel: Observed abundances (filled circles
with error bars) in comparison with the expected abundances in the
secondary star after having captured 17.5\% of the matter ejected  
within the solid angle subtended by the secondary from a metal-poor
($Z=0.001$) 40~\Msun spherically symmetric supernova explosion ($M_{\rm
He} = 15.8$ \Msuno) with $E_K = 10^{51}$ erg for two different
mass-cuts, $M_{\rm cut} = 2.51$~\Msun (solid line with open circles)
and $M_{\rm cut} = 3.01$~\Msun (dashed-dotted line with open circles).
The initial abundances of the secondary star were adopted for the
average abundances of halo stars with [Fe/H] $ = -1.4 \pm 0.2$, 
and the initial orbital distance was $a_{c,i} \approx 6$ \Rsuno. Right
panel: same as left panel but for a spherically symmetric
hypernova explosion ($E_K = 30 \times 10^{51}$ erg) for two different
mass-cuts, $M_{\rm cut} = 4.05$~\Msun (solid line with open circles)
and $M_{\rm cut} = 5.03$~\Msun (dashed-dotted line with open circles).}}
\label{fig9}
\end{figure*}

\subsection{Spherical Explosion}

Gelino et al. (2006) derived a current black hole mass of $M_{{\rm
BH},f}= 8.53\pm 0.60$~\Msun and secondary mass of $M_{2,f}= 0.37\pm
0.03$~\Msuno. Using near-UV spectroscopic observations of the 
accretion disk, Haswell et al. (2002) suggested that the material
accreted onto the compact object is substantially CNO processed,
indicating that the initial mass of the secondary star could have
been as high as $\sim 1.5$~\Msuno. Hereafter we will adopt an initial
secondary mass of $M_{2,i} = 1$~\Msun and a black hole mass of $M_{{\rm
BH},i} = 8$~\Msuno. 

A binary system such as \mbox{XTE J1118+480} will
survive a spherical SN explosion if the ejected mass $\Delta
M=M_{\rm He}-M_{{\rm BH},i} \le (M_{\rm He}+M_{2,i})/2$ (Hills 1983).
This implies a mass of the He core before the SN explosion of $M_{\rm
He} \le 17$ \Msuno. Using the expressions given by Portegies Zwart et
al. (1997a, and references therein), we inferred a He core radius of
$R_{\rm He} \approx 2-3$ \Rsun for He core masses in the range $M_{\rm He}
\approx 8.5-17$ \Msuno. We will assume that the post-SN orbital
separation after tidal circularization of the orbit is in the range
$a_{c,i} \approx 4-6$ \Rsuno, since the secondary must have experienced
mass and angular momentum losses during the binary evolution until
reaching its present configuration, with $a_{c,f} \approx 2.67$ \Rsuno.

Assuming a pre-SN circular orbit and an instantaneous spherically
symmetric ejection (that is, shorter than the orbital period), one can
estimate the pre-SN orbital separation, $a_0$, using the relations
given by van den Heuvel \& Habets (1984): $a_0=a_{c,i}\mu_f$, where
$\mu_f=(M_{{\rm BH},i}+M_{2,i})/(M_{\rm He}+M_{2,i})$. We find $a_0
\approx 3$--5, essentially depending on the adopted values of $M_{\rm He}$ 
and $a_{c,i}$. At the time of the SN explosion ($\sim$5--6 Myr; Brunish \&
Truran 1982), a 1~\Msun secondary star, still in its pre-main-sequence
evolution, has a radius $R_{2,i} \approx 1.3$
\Rsun and a convective zone of mass $M_{\rm cz} \approx 0.652$ \Msun
(D'Antona \& Mazzitelli 1994). Thus, the amount of
mass deposited on the secondary can be estimated as $m_{\rm
cap}=\Delta M (\pi R_{2,i}^2/4 \pi a_02)f_{\rm cap}$ \Msuno, where
$f_{\rm cap}$ is the fraction of mass, ejected within the solid angle
subtended by the secondary star, that is eventually captured. We
assume that the captured mass, $m_{\rm cap}$, is efficiently mixed with
the mass of the convective zone, $M_{\rm cz}$.

We compute the expected abundances in the atmosphere of the secondary
star after the pollution from the progenitor of the compact
object as in Gonz\'alez Hern\'andez et al. (2004). We used 40~\Msun
spherically symmetric core-collapse explosion models ($M_{\rm 
He} \approx 15.1$--16.1~\Msuno) for different metallicities ($Z=$ 0, 0.001,
0.004, 0.02) and explosion energies (Umeda \& Nomoto, 2002, 2005;
Tominaga, Umeda \& Nomoto 2007). These models imply $\Delta M \approx 
7$--8~\Msun and need small capture efficiencies of $f_{\rm
cap} \la 0.1$ (i.e., 10\%) to increase significantly the metal content of
the secondary star. On the other hand, the use of 30~\Msun models
($M_{\rm He} \approx 8.5$--11.2~\Msuno) would require $f_{\rm cap} \approx
0.9$--1. These models would also provide a different mass fraction of
each element at each value of the mass-cut (the mass that initially 
collapsed forming the compact remnant). For more details in the models, 
see Tominaga, Umeda \& Nomoto (2007).

The explosion energy, $E_K = 1 \times 10^{51}$ erg and $E_K = 
(20-30) \times 10^{51}$ erg for the supernova (SN) and hypernova (HN)
models (respectively), is deposited instantaneously in the central
region of the progenitor core to generate a strong shock wave. The
subsequent propagation of the shock wave is followed through a
hydrodynamic code (Umeda \& Nomoto 2002, and references therein). In
our simple model, we have assumed different mass-cuts, fallback
masses, and a mixing factor of 1 which assumes that all fallback
matter is well mixed with the ejecta. The amount of fallback, $M_{\rm
fall}$, is the difference between the final remnant mass, $M_{{\rm
BH},i}$, and the initial remnant mass of the explosion, $M_{\rm cut}$.
We should recall here the ejected mass, $\Delta M$, which is equal to
$M_{\rm He}-M_{{\rm BH},i}$, where $M_{\rm He}$ is the mass of the He
core.

\begin{figure*}[ht!]
\centering
\includegraphics[width=8.cm,angle=0]{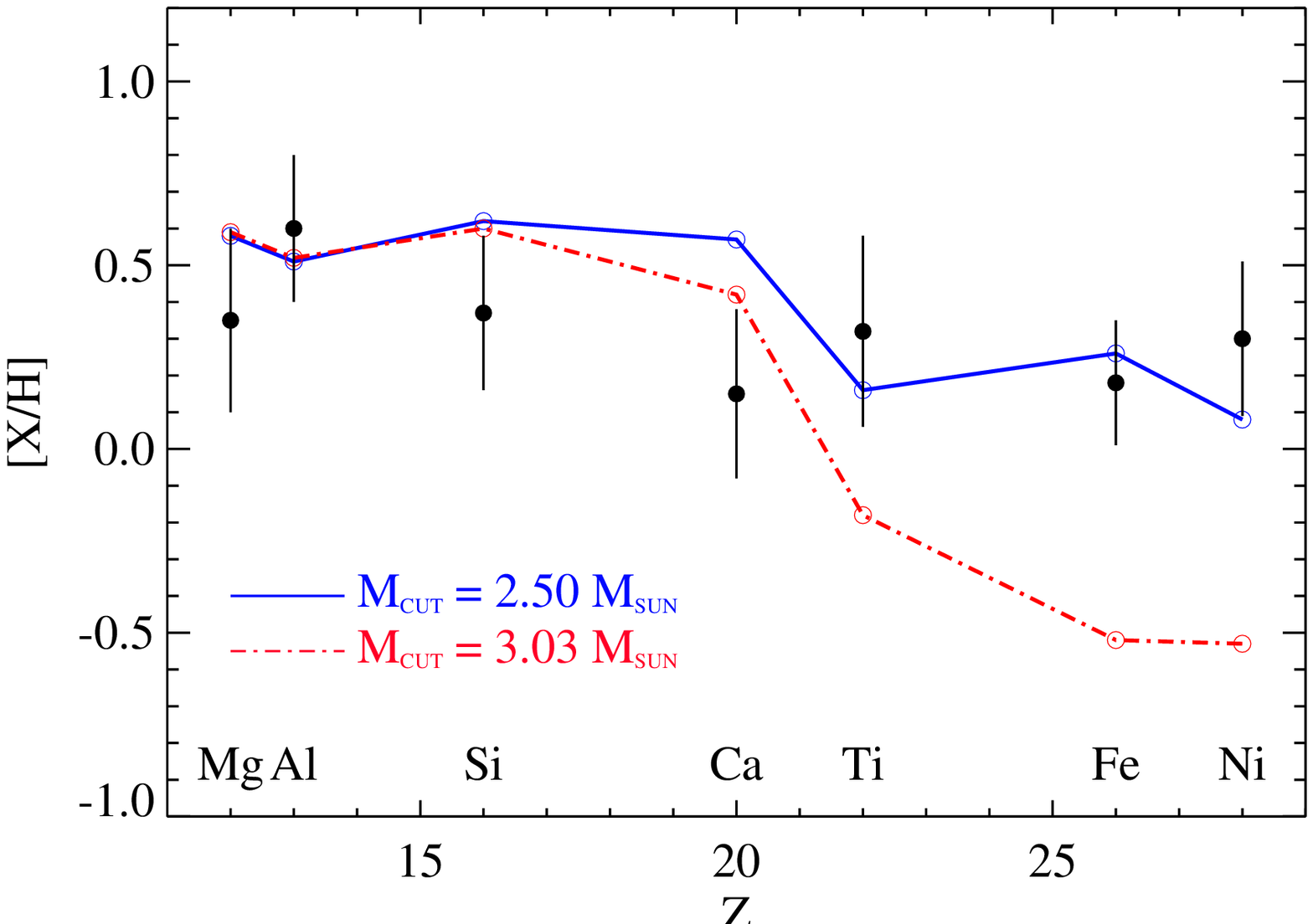}
\includegraphics[width=8.cm,angle=0]{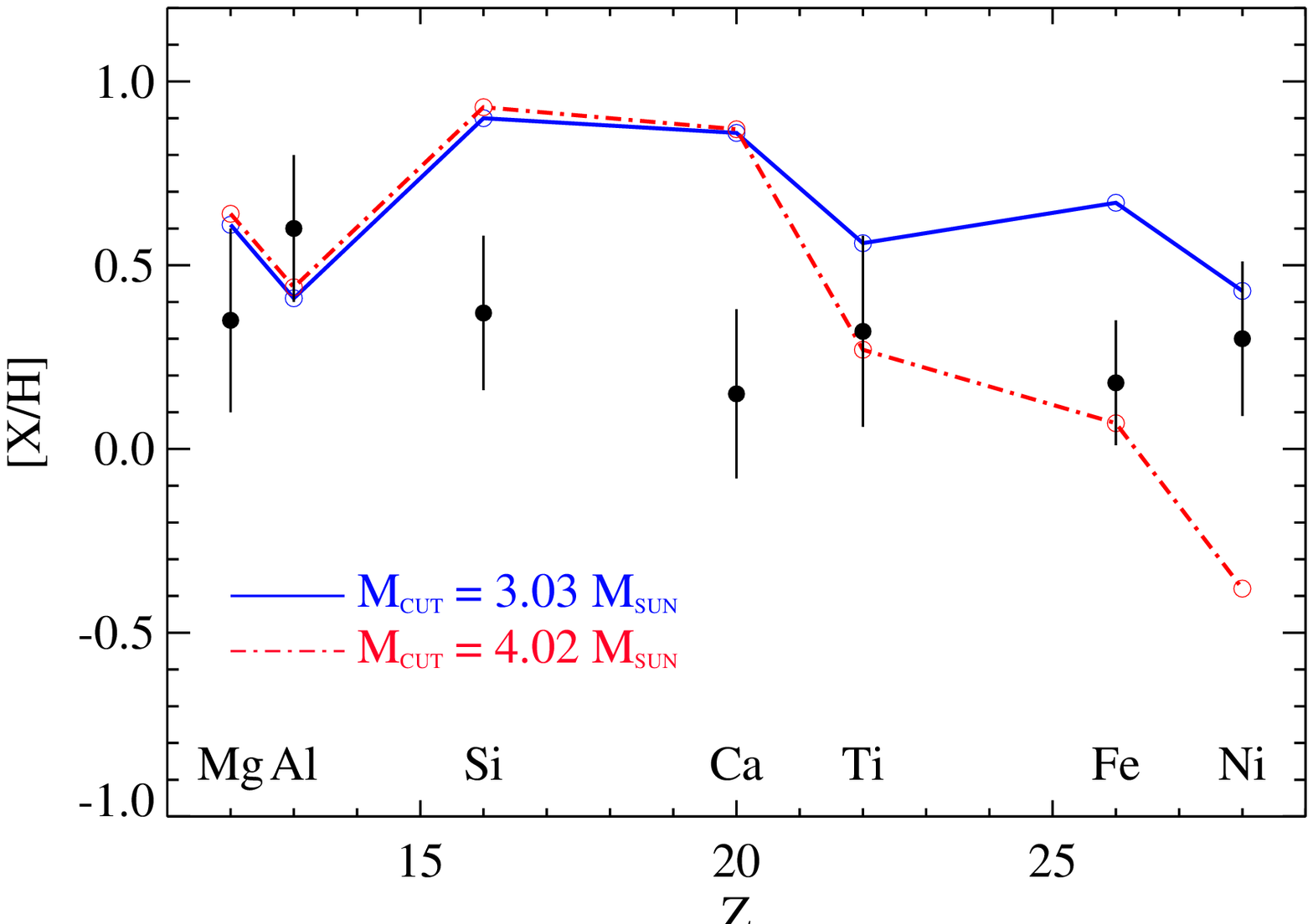}
\caption{\footnotesize{Left panel: Observed abundances (filled circles
with error bars) in comparison with the expected abundances in the
secondary star after having captured the 15.5\% of the matter ejected  
within the solid angle subtended by the secondary from a metal-poor
($Z = 0.004$) 40~\Msun spherically symmetric supernova explosion ($M_{\rm
He} = 15.8$ \Msuno) with $E_K = 10^{51}$ erg for two different
mass-cuts, $M_{\rm cut} = 2.50$~\Msun (solid line with open circles)
and $M_{\rm cut} = 3.03$~\Msun (dashed-dotted line with open circles).
The initial abundances of the secondary star were adopted for the
average abundances of thick-disk stars with [Fe/H] $= -0.7 \pm 0.2$,
and the initial orbital distance was $a_{c,i} \approx 6$~\Rsuno.
Right panel: same as left panel but for a spherically symmetric 
hypernova explosion ($E_K = 30 \times 10^{51}$ erg) for two different
mass-cuts, $M_{\rm cut} = 3.03$~\Msun (solid line with open circles)
and $M_{\rm cut} = 4.02$~\Msun (dashed-dotted line with open circles).}}
\label{fig10}
\end{figure*}

We use SN/HN models to provide us with the yields of the 
explosion before radiative decay of element species. We then 
compute the integrated, decayed yields of the ejecta by adopting 
a mass-cut and by mixing all the material above the mass-cut. 
Finally, we calculate the composition of the matter captured by
the secondary star, and we mix it with the material of its convective
envelope.

In Fig.~\ref{fig8} we show the expected abundances of the secondary 
star after contamination from the nucleosynthetic products of the SN
explosion ($E_K = 1 \times 10^{51}$ erg) of $M_{\rm He} \approx 
15$--16~\Msun progenitor stars. The initial abundances of the 
secondary star have been estimated from the average abundances 
of halo stars with a metallicity of [Fe/H] $\approx -2.2$ (from 
Cayrel et al. 2004) and [Fe/H] $\approx -1.4$ (from
Jonsell et al. 2005), thick-disk stars with [Fe/H] $\approx -0.7$ (from
Jonsell et al. 2005), and thin-disk stars with [Fe/H] $\approx 0$ and
[Fe/H] $\approx 0.18$ (from Gilli et al. 2006). For each simulation at 
each metallicity, we have fixed the $f_{\rm cap}$ factor at a value that
allows us to approximately match the observed aluminum abundance. 

We should remark that for a given model at a given metallicity, once the
capture efficiency is fixed, the aluminum (Al) abundance in the secondary
star hardly depends on the mass-cut. Thus, the fact that the Al
abundance in Fig.~\ref{fig8} is almost constant with metallicity is
because we have adopted different $f_{\rm cap}$ factors for different
metallicities. These $f_{\rm cap}$ factors are higher for lower
metallicities since greater amounts of captured material are needed to
achieve the observed Al abundance in the secondary star. For other
figures in this paper, the $f_{\rm cap}$ factor was changed until an
abundance of [Al/H]$ \approx 0.5$ dex was obtained, compatible with
the observed abundance within the uncertainties. 

\begin{deluxetable}{lcccccccccccc}
\centering
\tabletypesize{\scriptsize}
\tablecolumns{7}
\tablecaption{Metal-Poor Supernova/Hypernova Explosion Models in \mbox{XTE
J1118+480}}
\tablewidth{0pt}
\tablehead{ & & & \multicolumn{4}{c}{${\rm [X/H] \: EXPECTED}$\tablenotemark{d}}\\
\cline{4-7}\\
 & & & \multicolumn{2}{c}{Supernova} &
\multicolumn{2}{c}{Hypernova} \\
\cline{4-5} \cline{6-7} \\
ELEMENT & ${\rm [X/H]\:\rm OBSERVED}$\tablenotemark{a} & ${\rm [X/H]}_{0}$\tablenotemark{b} &
$M_{\rm cut, low}$\tablenotemark{c} & $M_{\rm cut, up}$ & $M_{\rm cut, low}$ & $M_{\rm cut, up}$}
\startdata
 & & & \multicolumn{4}{c}{40 \Msun explosion model of $Z=0.001$} \\
\noalign{\smallskip}
\tableline
\noalign{\smallskip}
Mg & 0.35 & -1.08 &  0.83 &  0.85 &  0.75 &  0.79  \\
Al & 0.60 & -1.27 &  0.52 &  0.53 &  0.39 &  0.42  \\
Si & 0.37 & -1.13 &  0.88 &  0.89 &  1.08 &  1.09  \\
Ca & 0.15 & -1.12 &  0.74 &  0.75 &  1.09 &  1.01  \\
Ti & 0.32 & -1.15 &  0.14 &  0.11 &  0.54 &  0.12  \\
Fe & 0.18 & -1.46 &  0.47 &  0.14 &  0.63 & -0.22  \\
Ni & 0.30 & -1.48 &  1.09 & -0.52 &  0.14 & -0.81  \\
O  & \nodata & -0.76 &  0.75 &  0.77 &  0.69 &  0.73  \\
C  & \nodata & -0.86 & -0.54 & -0.53 & -0.60 & -0.58  \\
\noalign{\smallskip}
\tableline
\noalign{\smallskip}
 & & & \multicolumn{4}{c}{40 \Msun explosion model with $Z=0.004$}  \\
\noalign{\smallskip}
\tableline
\noalign{\smallskip}
Mg & 0.35 & -0.41 &  0.58 &  0.59 &  0.61 &  0.64 \\
Al & 0.60 & -0.46 &  0.51 &  0.52 &  0.41 &  0.44 \\
Si & 0.37 & -0.50 &  0.62 &  0.60 &  0.90 &  0.93 \\
Ca & 0.15 & -0.54 &  0.57 &  0.42 &  0.86 &  0.87 \\
Ti & 0.32 & -0.50 &  0.16 & -0.18 &  0.56 &  0.27 \\
Fe & 0.18 & -0.75 &  0.26 & -0.52 &  0.67 &  0.07 \\
Ni & 0.30 & -0.73 &  0.08 & -0.53 &  0.43 & -0.38 \\
O  & \nodata & -0.26 &  0.71 &  0.73 &  0.67 &  0.69  \\
C  & \nodata & -0.35 &  0.10 &  0.11 &  0.06 &  0.08  \\
\noalign{\smallskip}
\tableline
\noalign{\smallskip}
 & & & \multicolumn{4}{c}{30 \Msun explosion model with $Z=0.004$} \\
\noalign{\smallskip}
\tableline
\noalign{\smallskip}
Mg & 0.35 & -0.41 &  0.60 &  0.62 &  0.57 &  0.62  \\
Al & 0.60 & -0.46 &  0.52 &  0.54 &  0.40 &  0.45  \\
Si & 0.37 & -0.50 &  0.77 &  0.54 &  1.02 &  0.95  \\
Ca & 0.15 & -0.54 &  0.72 &  0.17 &  1.03 &  0.82  \\
Ti & 0.32 & -0.50 &  0.18 & -0.41 &  0.66 &  0.01  \\
Fe & 0.18 & -0.75 &  0.14 & -0.71 &  0.71 & -0.32  \\
Ni & 0.30 & -0.73 & -0.43 & -0.54 &  0.40 & -0.40  \\
O  & \nodata & -0.26 &  0.75 &  0.78 &  0.69 &  0.74  \\
C  & \nodata & -0.35 &  0.02 &  0.03 & -0.04 & -0.01  \\
\enddata
\tablenotetext{a}{Observed abundances of the
secondary star in \mbox{XTE J1118+480}.}

\tablenotetext{b}{Initial abundances assumed for
the secondary star in \mbox{XTE J1118+480}, see text. The initial
C and O abundances of the thick-disk model (${\rm [X/H]}_{0}=-0.7$)
were adopted from Ecuvillon et al. (2004, 2006).}

\tablenotetext{c}{$M_{\rm cut, low}$ and $M_{\rm cut, up}$ are the
lower and upper mass-cut adopted in the model computation. See the
exact value in the captions of Figs.~9-11.} 

\tablenotetext{d}{Expected abundances of the secondary star.}

\tablecomments{Expected abundances in the secondary atmosphere
contaminated with nucleosynthetic products of metal-poor explosion
models for two different explosion energies and mass-cuts, presented
in Figs.~9-11.} 

\label{tbl5}
\end{deluxetable}

   From Fig.~\ref{fig8} one can see that while Mg and Al remain 
significantly enhanced (above solar abundances) for all mass-cuts at 
all metallicities, Si, Ca, Ti, Fe, and Ni are quite sensitive to the
mass-cut of the model. Thus, for metallicities below $-1.4$ dex, only
$M_{\rm cut} \la 3$~\Msun is able to enhance sufficiently the Ti, Fe, and
Ni, whereas Ca and Si remain quite overabundant for $M_{\rm cut} \la 
4$~\Msun and 5~\Msun (respectively) at all metallicities. We should note
that for metallicities [Fe/H] $\la -1.4$ dex (i.e., $Z \la 0.001$), the
expected abundances in the secondary star do not depend on the initial
abundances but on the SN yields. Thus, we found different expected
abundances of the secondary between [Fe/H] $\approx -1.4$ dex and
[Fe/H] $\approx -2.2$ dex because we used the SN $Z = 0.001$ model and 
the SN $Z = 0$ model, respectively. 

\subsection{Formation in the Halo}

The kinematics of the system, at least $U$ and $V$, resemble those
of halo stars, and therefore a significant kick during the black
hole formation process appears unnecessary. Thus, a spherically
symmetric SN explosion would provide the desired kick velocity to
match the current velocity components of the system from initial
velocities similar to those of halo stars. However, due to the 
extremely low metallicities of halo stars, it is required that the 
secondary captured enough matter from the ejecta to reach the current 
abundances. 

\begin{figure*}[ht!]
\centering
\includegraphics[width=8.cm,angle=0]{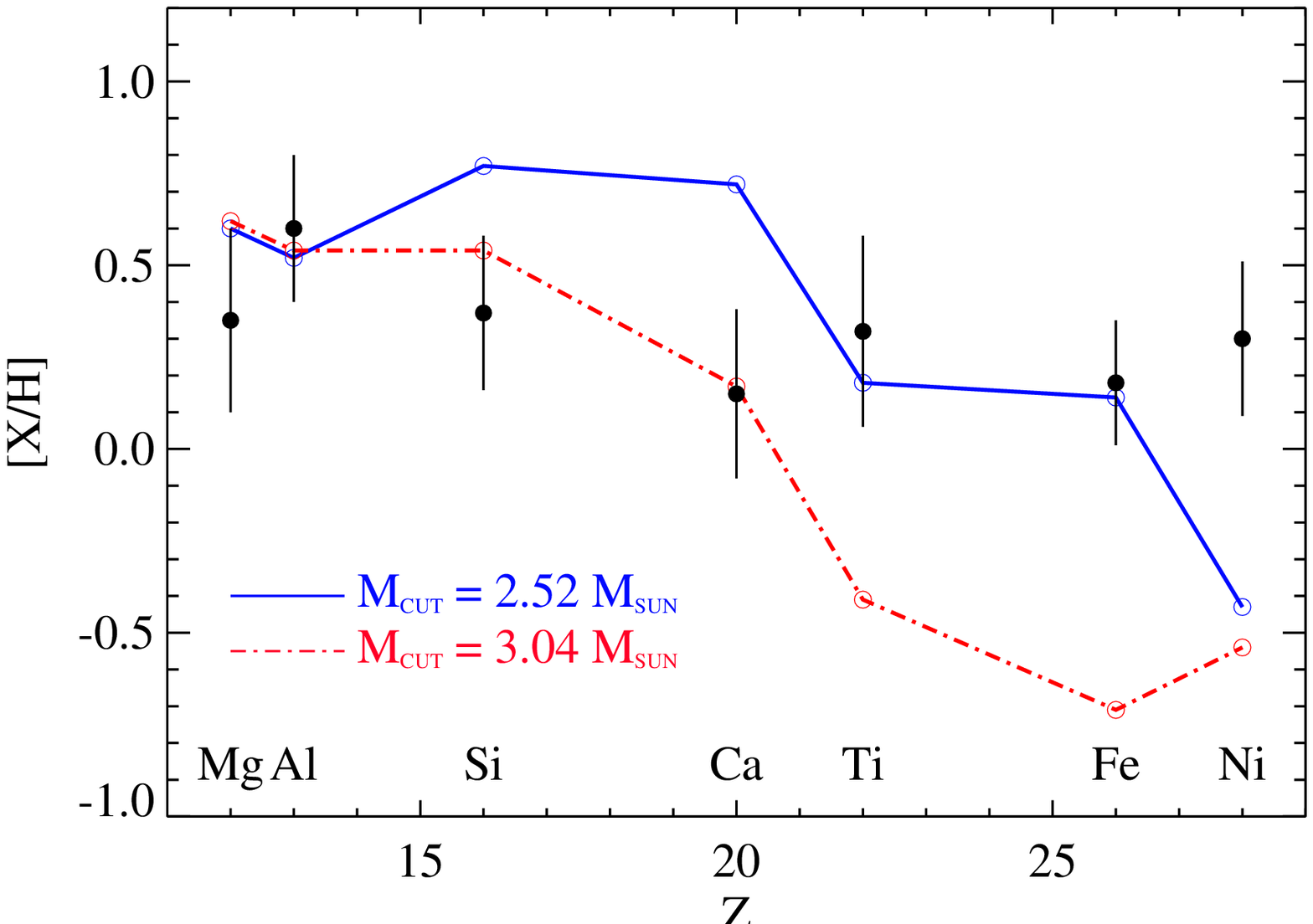}
\includegraphics[width=8.cm,angle=0]{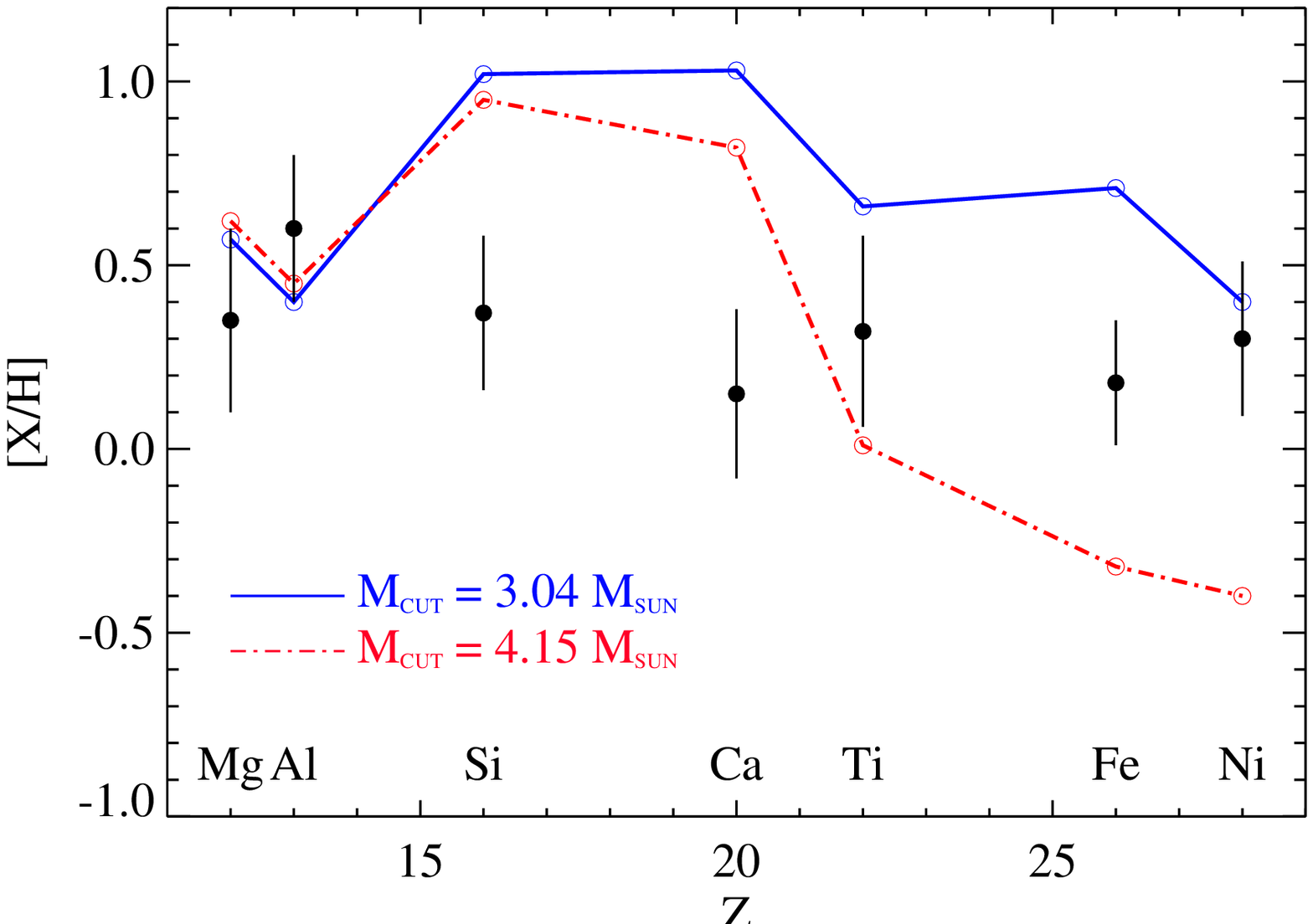}
\caption{\footnotesize{Left panel: Observed abundances (filled circles
with error bars) in comparison with the expected abundances in the
secondary star after having captured 90\% of the matter ejected  
within the solid angle subtended by the secondary from a metal-poor
($Z = 0.004$) 30~\Msun spherically symmetric supernova explosion 
($M_{\rm He} = 11$ \Msuno) with $E_K = 10^{51}$ erg for two different
mass-cuts, $M_{\rm cut} = 2.52$~\Msun (solid line with open circles)
and $M_{\rm cut} = 3.04$~\Msun (dashed-dotted line with open circles).
The initial abundances of the secondary star were adopted for the
average abundances of the thick-disk stars with [Fe/H] $= -0.7 \pm 0.2$,
and the initial orbital distance was $a_{c,i} \approx 6$ \Rsuno.
Right panel: same as left panel but for a spherically symmetric
hypernova explosion ($E_K = 20 \times 10^{51}$ erg) for two different
mass-cuts, $M_{\rm cut} = 3.04$~\Msun (solid line with open circles)
and $M_{\rm cut} = 4.15$~\Msun (dashed-dotted line with open circles).}}
\label{fig11}
\end{figure*}

\begin{figure*}[ht!]
\centering
\includegraphics[width=8.cm,angle=0]{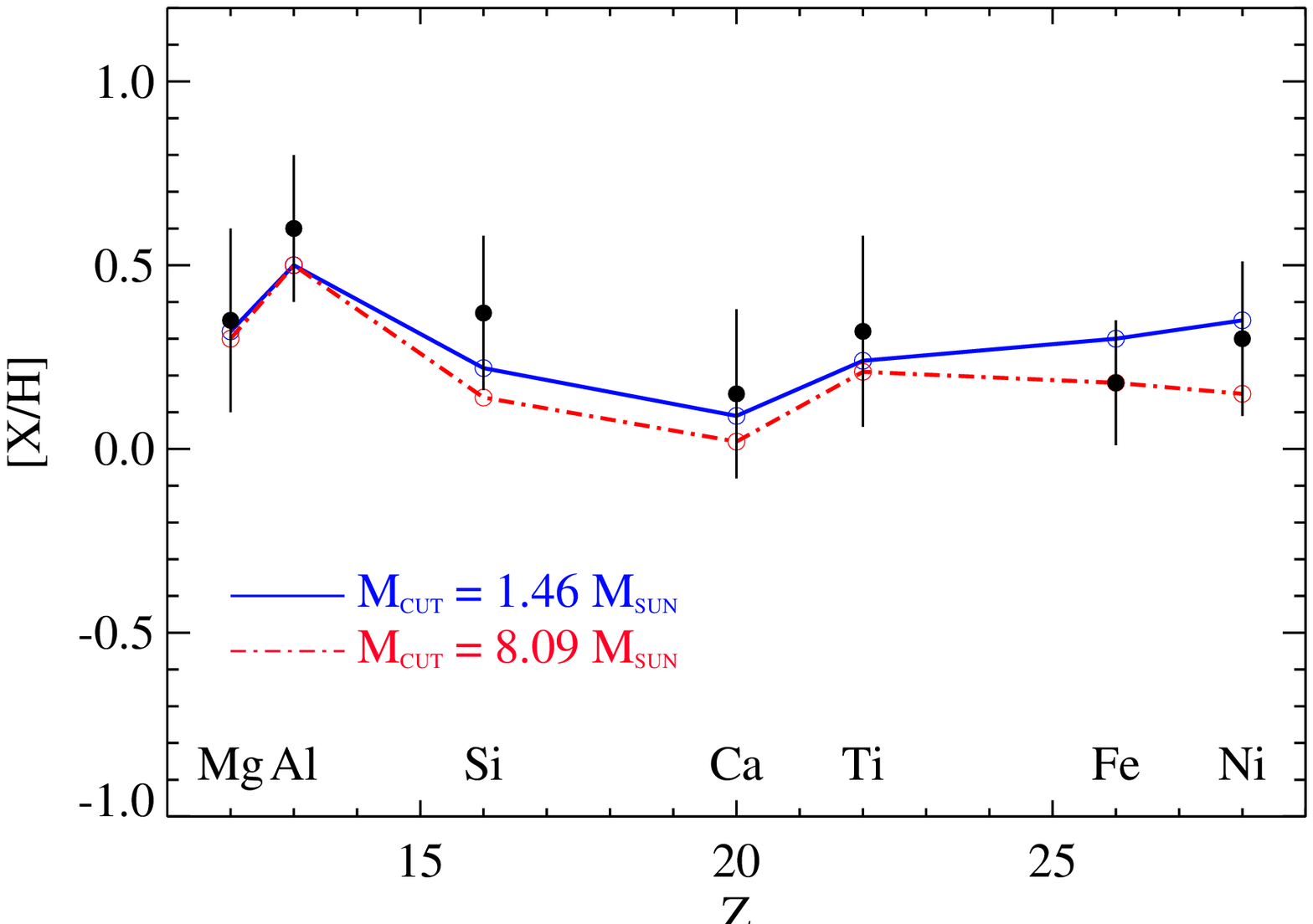}
\includegraphics[width=8.cm,angle=0]{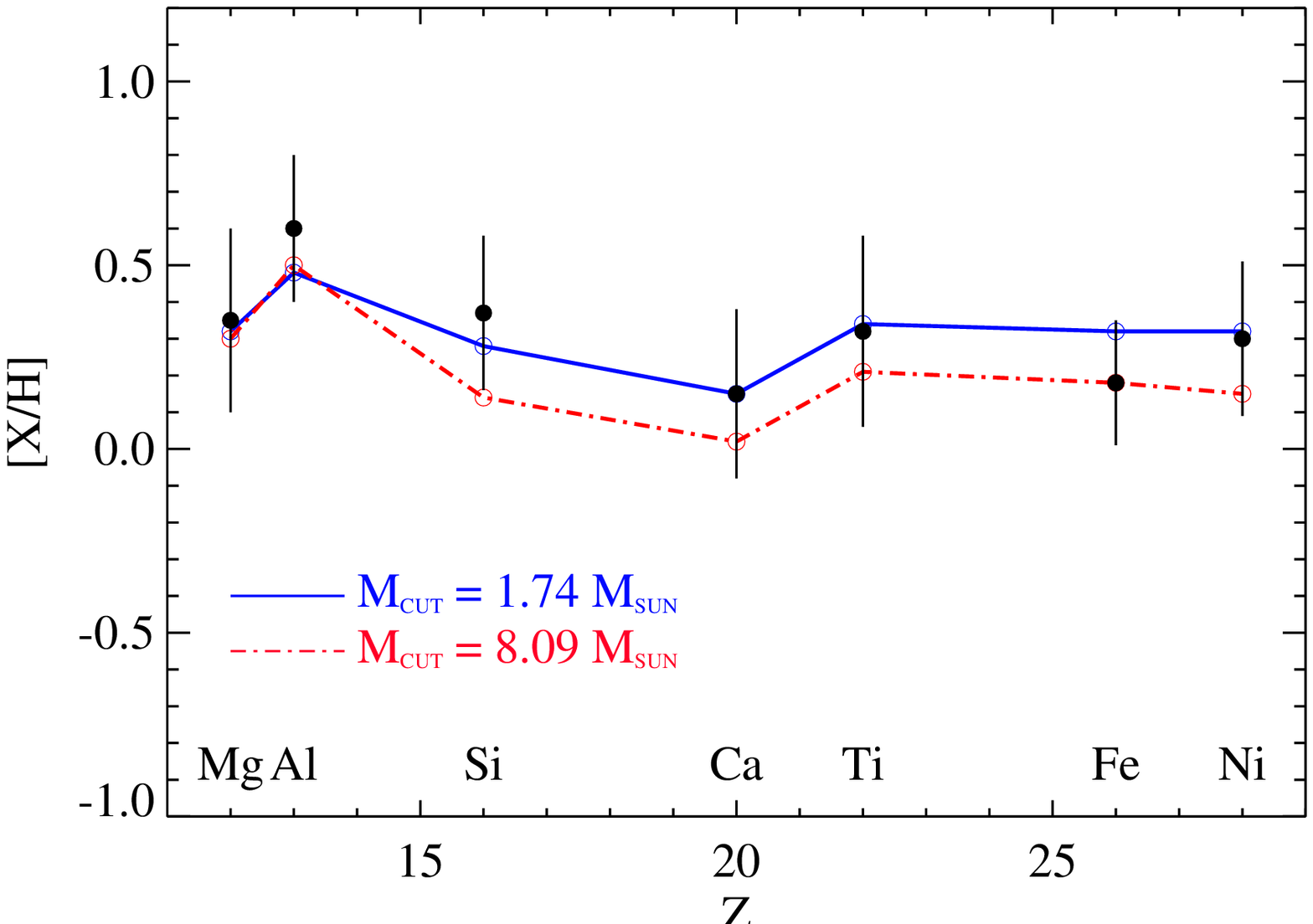}
\caption{\footnotesize{Left panel: Observed abundances (filled circles
with error bars) in comparison with the expected abundances in the
secondary star after having captured the 5\% of the matter ejected  
within the solid angle subtended by the secondary from a metal-poor
($Z = 0.02$) 40~\Msun spherically symmetric supernova explosion ($M_{\rm
He} = 15.1$~\Msuno) with $E_K = 10^{51}$ erg for two different
mass-cuts, $M_{\rm cut} = 1.46$~\Msun (solid line with open circles)
and $M_{\rm cut} = 8.09$~\Msun (dashed-dotted line with open circles).
The initial abundances of the secondary star were adopted for the
average abundances of thin-disk stars with [Fe/H] $=0.0 \pm 0.2$.
Right panel: same as left panel but for a spherically symmetric 
hypernova explosion ($E_K = 30 \times 10^{51}$ erg) for two different
mass-cuts, $M_{\rm cut} = 1.74$~\Msun (solid line with open circles)
and $M_{\rm cut} = 8.09$~\Msun (dashed-dotted line with open circles).}}
\label{fig12}
\end{figure*}

In Fig.~\ref{fig9} we compare the observed abundances
with the expected abundances in the secondary star for two different
explosion energies and mass-cuts using a metal-poor 40~\Msun
progenitor model. In these simulations, the mass-cut is sampled
according to the mass bins given in the explosion models, from
$\sim 1.5$~\Msun to 8~\Msun in steps of $\sim 0.5$ and 1~\Msun. The
adopted mass-cuts shown in Fig.~\ref{fig9} were selected from those
which provided better fits to the observed abundances. The initial
abundances of the secondary have been estimated from the average 
abundances of halo stars with [Fe/H] $\approx -1.4$ from Jonsell et 
al. (2005). The value of $f_{\rm cap}$ has
been adjusted until the observed aluminum abundance was roughly
reproduced, providing $f_{\rm cap} \approx 0.17$. The parameters of
the explosion models used in Fig.~\ref{fig9}~--~\ref{fig13} are 
given in Table~\ref{tbl4}. In Table~\ref{tbl5} we show the
expected abundances of the secondary star after contamination from
metal-poor SN/HN explosion models.

With a supernova model, shown in the left panel of Fig.~\ref{fig9}, 
it is not possible to recover the observed abundances because 
these models produce too much Mg and Ca regardless of the Fe 
abundance obtained; in addition, the Ni abundance is
strongly dependent on the mass-cut. The hypernova model, displayed in
the right panel, makes it even more difficult to fit the observed
abundances since this model creates too much Ca relative to Fe and
Ni. As inferred from Fig.~\ref{fig8}, a model with initial abundances 
at [Fe/H] $\approx -2.2$ (Cayrel et al. 2004) would also not be
successful in reproducing the observed abundances due to the
tremendous and different sensitivity of each element to the mass-cut. 

In conclusion, neither of these very metal-poor models  
is able to reproduce the observed abundances. We also try to fit
the observed abundances using a metal-poor 30~\Msun explosion model,
but the agreement is even worse than for 40~\Msun models, again due 
to the strong sensitivity of the each element abundance to the
mass-cut. The comparison of the observed abundances with SN yields
allows us to rule out a Galactic halo origin for this black hole
binary. 

\subsection{Formation in the Thick Disk}

\begin{deluxetable}{lcccccccccccc}
\centering
\tabletypesize{\scriptsize}
\tablecolumns{7}
\tablecaption{Metal-Rich Supernova/Hypernova Explosion Models in \mbox{XTE
J1118+480}}
\tablewidth{0pt}
\tablehead{ & & & \multicolumn{4}{c}{${\rm [X/H] \: EXPECTED}$\tablenotemark{d}}\\
\cline{4-7}\\
ELEMENT & ${\rm [X/H]\:\rm OBSERVED}$\tablenotemark{a} & ${\rm [X/H]}_{0}$\tablenotemark{b} &
$M_{\rm cut, low}$\tablenotemark{c} & $M_{\rm cut, up}$ & $M_{\rm cut, low}$ & $M_{\rm cut, up}$}
\startdata
 & & & \multicolumn{4}{c}{Spherical explosion model of $Z=0.02$} \\
\noalign{\smallskip}
\tableline
\noalign{\smallskip}
 & & & \multicolumn{2}{c}{Supernova} &
\multicolumn{2}{c}{Hypernova} \\
\noalign{\smallskip}
\cline{4-5} \cline{6-7} \\
\noalign{\smallskip}
Mg & 0.35 & 0.17 &  0.32 &  0.30 &  0.32 &  0.30  \\
Al & 0.60 & 0.29 &  0.50 &  0.50 &  0.48 &  0.50  \\
Si & 0.37 & 0.12 &  0.22 &  0.14 &  0.28 &  0.14  \\
Ca & 0.15 & 0.02 &  0.09 &  0.02 &  0.15 &  0.02  \\
Ti & 0.32 & 0.20 &  0.24 &  0.21 &  0.34 &  0.21  \\
Fe & 0.18 & 0.18 &  0.30 &  0.18 &  0.32 &  0.18  \\
Ni & 0.30 & 0.13 &  0.35 &  0.15 &  0.32 &  0.15  \\
O  & \nodata & 0.18 &  0.33 &  0.33 &  0.33 &  0.33  \\
C  & \nodata & 0.11 &  0.16 &  0.16 &  0.15 &  0.16  \\
\noalign{\smallskip}
\tableline
\noalign{\smallskip}
 & & & \multicolumn{4}{c}{Aspherical explosion model with $Z=0.02$}  \\
\noalign{\smallskip}
\tableline
\noalign{\smallskip}
 & & & \multicolumn{2}{c}{Angle\tablenotemark{e} $=0^\circ-15^\circ$}
 & \multicolumn{2}{c}{Angle $=0^\circ-90^\circ$} \\
\noalign{\smallskip}
\cline{4-5} \cline{6-7} \\
\noalign{\smallskip}
Mg & 0.35 & 0.17 &  0.34 &  0.32 &  0.34 &  0.35 \\
Al & 0.60 & 0.29 &  0.51 &  0.48 &  0.50 &  0.51 \\
Si & 0.37 & 0.12 &  0.24 &  0.18 &  0.35 &  0.26 \\
Ca & 0.15 & 0.02 &  0.05 &  0.02 &  0.22 &  0.11 \\
Ti & 0.32 & 0.20 &  0.21 &  0.20 &  0.41 &  0.35 \\
Fe & 0.18 & 0.18 &  0.18 &  0.18 &  0.28 &  0.22 \\
Ni & 0.30 & 0.13 &  0.13 &  0.13 &  0.34 &  0.23 \\
O  & \nodata & 0.18 &  0.37 &  0.35 &  0.38 &  0.39  \\
C  & \nodata & 0.11 &  0.12 &  0.13 &  0.12 &  0.13  \\
\enddata
\tablenotetext{a}{Observed abundances of the
secondary star in \mbox{XTE J1118+480}.}

\tablenotetext{b}{Initial abundances assumed for
the secondary star in \mbox{XTE J1118+480}, see text. The initial
C and O abundances of the metal-rich models were adopted from
Ecuvillon et al. (2004, 2006).}

\tablenotetext{c}{$M_{\rm cut, low}$ and $M_{\rm cut, up}$ are the
lower and upper mass-cut adopted in the model computation. See the
exact value in the captions of Figs.~12 and 13.} 

\tablenotetext{d}{Expected abundances of the secondary star.}

\tablenotetext{e}{Angular range, measured from the equatorial plane,
in which all the ejected material in the explosion has been completely
mixed for each velocity point.} 

\tablecomments{Expected abundances in the secondary atmosphere
contaminated with nucleosynthetic products of metal-rich explosion
models for two different mass-cuts and symmetries, presented
in Figs.~12 and 13.} 

\label{tbl6}
\end{deluxetable}

The space-velocity components of the system are comparable to those of
thick-disk stars, and its present location 1.6~kpc above the Galactic
plane is slightly higher that the scale height of thick-disk stars
($\sim 0.8$--1.3 kpc; Reyl\'e \& Robin 2001; Chen 1997); thus, a 
strong kick during the SN explosion would not be required. A
spherically symmetric SN explosion of a 15.8~\Msun and 11~\Msun He core
provides an impulse of $\sim 60$ and $\sim20$ \kms, respectively. 
However, an enrichment from the typical abundances of thick-disk stars
would have been necessary. In the simulations, the initial abundances
were assumed to be the average values of thick-disk stars with
[Fe/H] $\approx -0.7$ from Jonsell et al. (2005). 

  In Fig.~\ref{fig10} we compare the expected abundances from a
40~\Msun explosion model for two energies and mass-cuts (see also
Table~\ref{tbl5}). The left panel shows the expected abundances
from a SN model which seems to better approach the observed
abundances than those of halo-like metallicities. As in the previous
figure, the $f_{\rm cap}$ factor was changed until an abundance of
[Al/H]$ \approx 0.5$ dex was obtained, compatible with the observed
abundance within the uncertainties. For the mass-cut of 
2.50~\Msuno, the model roughly fits the observed abundances except for
Ca, which appears enhanced in the model by a factor of 2.6. Better
agreement with Ca could be obtained if the $f_{\rm cap}$ factor is
lowered, but then Al, Ni, and Ti would not match their observed
abundances. If we increase the mass-cut up to 3.03~\Msuno, the Ca
abundance only decreases 0.15 dex whereas Ti, Fe, and Ni decrease by a
factor of 2, 6, and 4 (respectively), which makes the model unable to fit
the observed abundances. The hypernova case offers worse results since
Si and Ca are no longer reproduced, and Fe and Ni cannot be fitted at
the same time.

For this metallicity we have also explored the 30~\Msun explosion
model. The results are displayed in Fig.~\ref{fig11}. In the left panel
the SN model is compared with the observations. For the lower
mass-cut, at 2.52~\Msuno, the model provides too high Si and Ca
abundances and too low Ni abundance, whereas for a mass-cut of 
3.04~\Msuno, the observed Ti, Fe, and Ni abundances are too high in
comparison with the model predictions. In the hypernova case (right
panel of Fig.~\ref{fig11}), at these low mass-cuts the model produces
too large amounts of Si and Ca; in addition, Ti, Fe, and Ni are too 
sensitive to the location of the mass-cut.

Despite the fact that none of the explosion models explored is able to
fairly reproduce the observed abundances in the secondary star, all of
the models require vigorous mixing between the fallback matter and
the final ejected matter (Kifonidis et al.\ 2000). For instance, in
the 15.8~\Msun He core model, all of the material above the mass-cut
placed at 2.50~\Msun should be well mixed in order to convey heavy
elements like Fe and Ni to the outer layers of the explosion, which
might make more unlikely a Galactic thick-disk origin for
\mbox{XTE J1118+480}.

\subsection{Formation in the Thin Disk}

In this scenario, the system must have acquired an impulse during the
formation of the black hole that pushes it up from a Galactic plane
orbit to the current halo orbit. The system should have been
accelerated to a peculiar space velocity of $\sim 180$ {\kms}
(Gualandris et al.\ 2005) to reach its present location, requiring 
an asymmetric kick. It has been suggested that such kicks can be
imparted during the birth of nascent neutron stars, due to asymmetric
mass ejection and/or an asymmetry in the neutrino emission (Lai et
al.\ 2001, and references therein). 

Podsiadlowski et al. (2002)
proposed that the black hole in \mbox{Nova Sco 1994} could have formed
in a two-stage process where the initial collapse led to the
formation of a neutron star accompanied by a substantial kick and the
final mass of the compact remnant was achieved by matter that fell
back after the initial collapse. However, the black hole mass in 
\mbox{Nova Sco 1994} is estimated to be $\sim 5.4$~\Msun (Beer \&
Podsiadlowski 2002), while the black hole in \mbox{XTE J1118+480} has
a mass of $\sim 8$ \Msun which would require a fallback mass of
$\sim 6.6$ \Msun if we assume $\sim 1.4$~\Msun for a canonical
neutron star. This scenario might take place in the context of
collapsar models where the black hole would be formed in a mild
explosion and substantial fallback (up to $\sim 5$~\Msun is expected) as 
proposed by MacFadyen et al. (2001). Asymmetric mass ejection would
relax this requirement, providing enough impulse to the system to be
launched into its present orbit from the Galactic plane.

\begin{figure*}
\centering
\includegraphics[width=8.cm,angle=0]{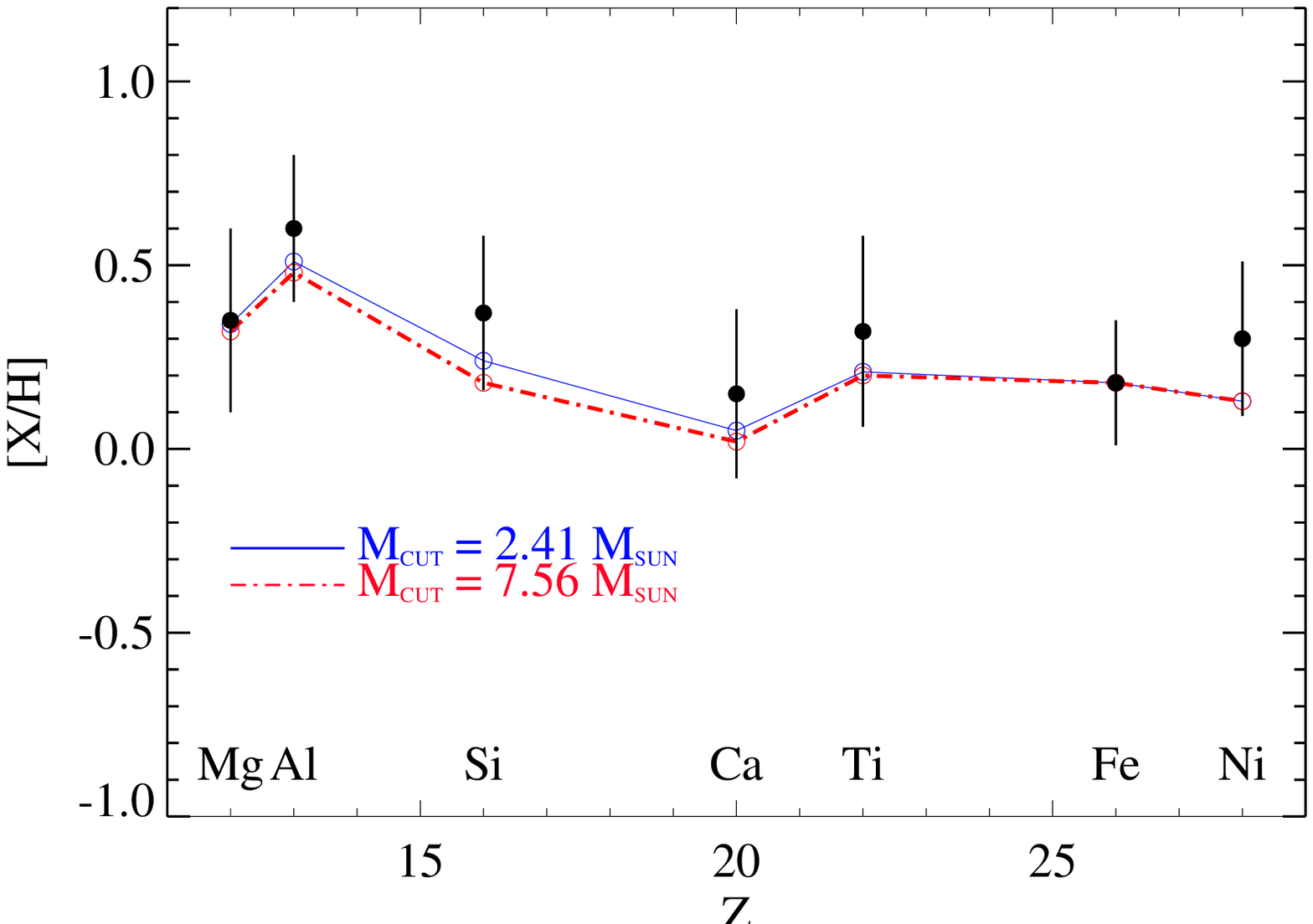}
\includegraphics[width=8.cm,angle=0]{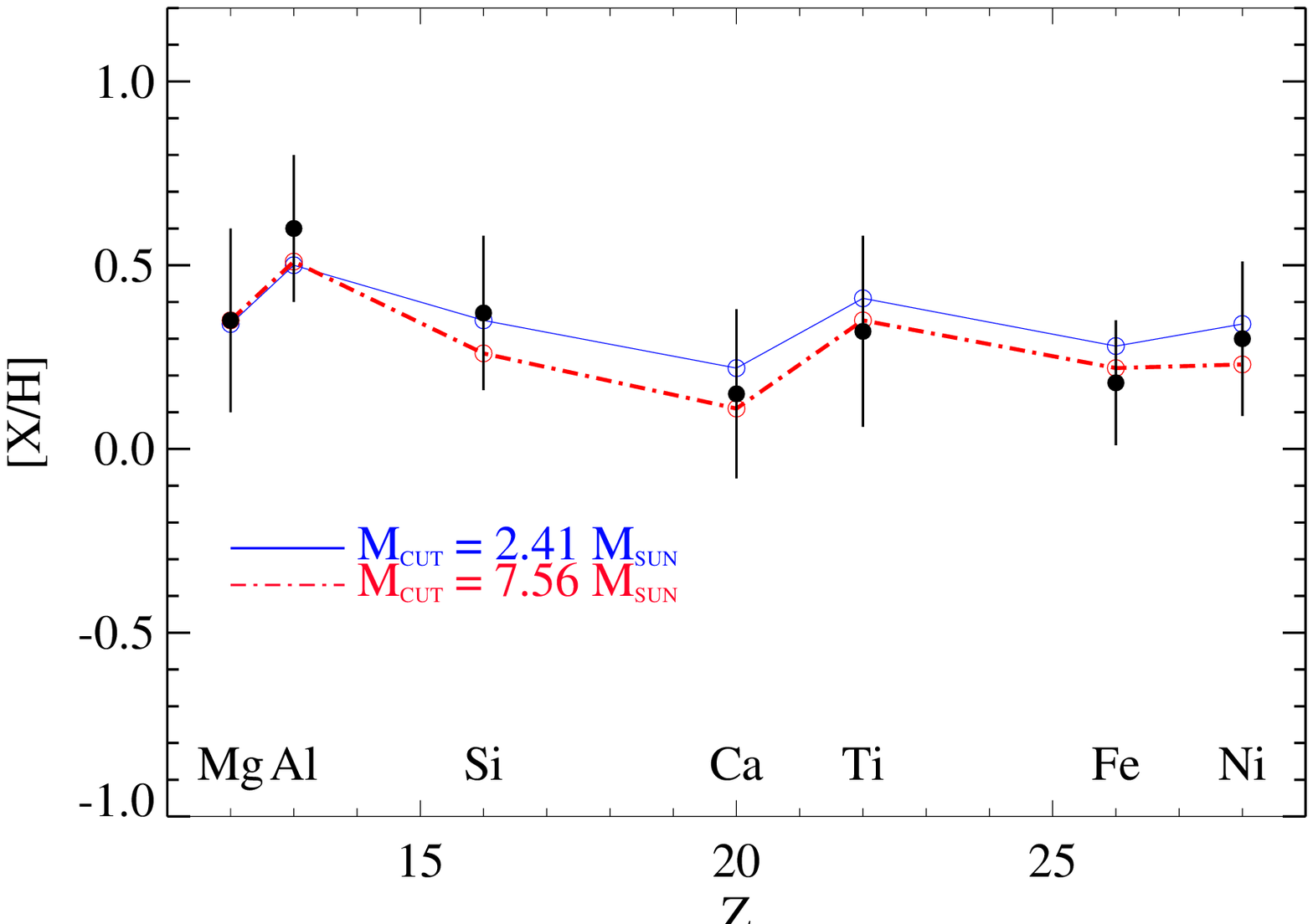}
\caption{\footnotesize{Left panel: Observed abundances (filled circles
with error bars) in comparison with the expected abundances in the
secondary star after having captured the 5\% of the matter ejected  
within the solid angle subtended by the secondary from a
non-spherically symmetric supernova explosion of $E_K =
10^{52}$ erg for two different mass-cuts, $M_{\rm cut} = 2.41$~\Msun
(solid line with open circles) and $M_{\rm cut} = 7.56$~\Msun
(dashed-dotted line with open circles). This model corresponds to the
matter ejected in the equatorial plane of the primary where we assumed
that the secondary star is located (for more details, see Gonz\'alez
Hern\'andez et al. 2005b). Right panel: same as left panel but in this model
we have assumed complete lateral mixing, where all the material within
given velocity bins is completely mixed. Two simulations
are shown for two different mass-cuts, $M_{\rm cut} = 2.41$~\Msun (solid
line with open circles) and $M_{\rm cut} = 7.56$~\Msun (dashed-dotted
line with open circles).}}  
\label{fig13}
\end{figure*}

In Fig.~\ref{fig12}, we compare the expected abundances from the
explosion of a 15.1~\Msun He core with the observed abundances of the
secondary star (see also Table~\ref{tbl6}). As in the previous
figures, the $f_{\rm cap}$ factor 
was changed until an abundance of [Al/H]$ \approx 0.5$ dex was obtained,
compatible with the observed abundance within the uncertainties. The
initial abundances were assumed to match the average values of disk
stars with similar iron content, which are provided in Table~\ref{tbl3}. 
For both the SN model (left panel) and the hypernova model 
(right panel), the observed element abundances can be reproduced for
all mass-cuts. This means that neither substantial fallback nor an
efficient mixing process is needed. For mass-cuts above 
$\sim 3$~\Msuno, very little Si, Ca, Ti, Fe, and Ni is ejected; 
thus, the expected abundances of the model essentially reflect
the initial abundances of the secondary star. In contrast, Mg and
Al are not sensitive to the mass-cut and are slightly enhanced due to
the capture of enriched material in the SN explosion. 

We also investigate a model with solar initial abundances for the
secondary star, i.e., ${\rm [X/H]}_{0}=0$, and the same explosion
model of solar metallicity ($Z=0.02$), which
would fit all the element abundances if the mass-cut is low enough
($M_{\rm cut} \la 3$~\Msuno) and mixing is so efficient that
significant amounts of elements which form in the inner layers of the
explosion (such as Ti, Ni, and Fe) are present in the ejecta.  

\subsection{Non-Spherical Explosion: Formation in the Thin Disk}

In this section, the thin-disk scenario is studied using
non-spherically symmetric explosion models from Maeda et al. (2002).
The chemical composition of the ejecta in a non-spherically symmetric
SN explosion is strongly dependent on direction. In particular, 
if we assume that the jet is collimated perpendicular to the orbital
plane of the binary, where the secondary star is located, elements
such as Ti, Fe, and Ni are mainly ejected in the jet direction, while
O, Mg, Al, Si, and S are preferentially ejected near the equatorial
plane of the helium star (Maeda et al.\ 2002). 

In Fig.~\ref{fig13} we compare the predicted abundances 
in the secondary star after pollution from an aspherical explosion 
model of a metal-rich progenitor having a 16~{\Msun} He core (see also
Table~\ref{tbl6}). The initial abundances of the secondary were
extracted from the average values of stars of the solar neighborhood
with similar iron content (see Table~\ref{tbl3}). The left panel
reflects the composition of the material ejected in the equatorial
plane while in the right panel we have considered complete lateral
mixing (Podsiadlowski et al.\ 2002) --- that is, the ejected matter is
completely mixed within each velocity bin. This mixing process is
done with the decayed yields of the model, and after that, all the
material above the mass-cut is mixed and used to calculate the  
composition of the ejected matter. The observed abundances are better
reproduced if complete lateral mixing is considered since this process
tends to enhance all of the element abundances at all mass-cuts. 
However, the equatorial model (left panel) also provides good fits to the
observed abundances. It should be noted that in this model, only the
material ejected in the equatorial plane is captured and therefore,
only Mg, Al, and Si are significantly enhanced with respect to the
initial abundances. A model with solar initial abundances for the
secondary star, i.e. ${\rm [X/H]}_{0}=0$, and the same explosion model of
solar metallicity ($Z=0.02$) was also inspected, providing the
same result except for the equatorial model which produces too low
abundances of Ti, Fe, and Ni in comparison with the observations.

Both aspherical explosion models with initial abundances equal to the
average values of thin-disk stars show little dependence on the
mass-cut, and since the equatorial model can reproduce the observed
abundances, extensive mixing processes are not required. In addition,
the non-spherically symmetric ejection of the mass in the explosion
could provide the kick that the system needs to change its orbit from
the Galactic plane to the current orbit. Therefore, the element
abundances in the secondary star and the kinematics of this system
strongly suggest that the binary system \mbox{XTE J1118+480} formed in
the Galactic disk (probably the thin disk) and it was then kicked
towards the Galactic halo, most probably by asymmetric mass ejection
in an asymmetric supernova/hypernova explosion that gave rise to the
black hole in this system.

The elements O and C, for which we provide the expected abundances
in Tables~\ref{tbl5} and~\ref{tbl6}, could be studied in future
investigations, probably from CO and OH bands in the near-IR. This
would help to recognize the operation of the CNO cycle on the surface
the secondary star, proposed by Haswell et al. (2002), and
possible processes of rotation-induced mixing during the evolution of
the massive star. 

\section{Conclusions}

We have presented Keck~II/ESI medium-resolution spectroscopy of the
black hole binary \mbox{XTE J1118+480}. The individual spectra of the
system allowed us to derive an orbital period of $P = 0.16995 \pm
0.00012$~d and a radial velocity semiamplitude of the secondary star
of $K_2 = 708.8 \pm 1.4$ \kmso. The implied updated mass function is
$f(M) = 6.27 \pm 0.04$~\Msun, consistent with (but more precise than)
previous values reported in the literature. Inspection of the
high-quality averaged spectrum of the secondary star provides a
rotational velocity of $v~\sin~i = 100^{+3}_{-11}$ \kmso, and hence a
binary mass ratio $q = 0.027 \pm 0.009$. The derived radial velocity,
$\gamma = 2.7 \pm 1.1$ \kmso, of the center of mass of the system
agrees, at the 3$\sigma$ level, with the results of previous studies.

We have performed a detailed chemical analysis of the secondary star.
We applied a technique that provides a determination of the stellar
parameters, taking into account any possible veiling from the
accretion disk. We find $T_{\mathrm{eff}} = 4700 \pm 100$~K, $\log 
[g/{\rm cm~s}^2] = 4.6 \pm 0.3$, $\mathrm{[Fe/H]} = 0.18 \pm 0.17$, 
and a disk veiling (defined as $F_{\rm  disk}/F_{\rm total}$) of
$\sim$40\% at 5000~{\AA}, decreasing toward longer wavelengths.  

We have provided further details on the abundances of Mg, Al,
Ca, Fe, Ni, and Li already reported by Gonz\'alez Hern\'andez et al.
(2006), and we determined new element abundances of Si and Ti. The chemical
abundances are typically higher than solar, and in some cases they 
are slightly enhanced (e.g., Mg, Al, and Si) in comparison with the
abundances of these elements in stars of the solar neighborhood 
having similar iron content.

The present location and kinematics of this binary system had suggested
that it could have originated in the Galactic halo. However, the
chemical abundances strongly indicate that the black hole formed as a
consequence of a supernova/hypernova explosion that occurred within
the binary system. This explosive event must have either provided a
kick to the system if it was formed in the thin disk or enriched
significantly the atmosphere of the secondary star if the system
formed in the thick disk or halo. 

We have explored a variety of supernova/hypernova explosion models for
different metallicities, He core masses, and geometries. We compared
the expected abundances in the secondary star after contamination from
nucleosynthetic products from different initial metallicities of the
secondary star ($-2.2 < \mathrm{[Fe/H]} < 0.2$), to investigate the formation 
region in the Galactic halo, thick disk, or thin disk. Metal-poor
explosion models ($Z = 0$ and $Z = 0.001$) were not able to fit the 
observed abundances since they produce inappropriate ratios between
$\alpha$-elements and iron-peak elements, and they are extremely
sensitive to the adopted mass-cut. This comparison probably rules out 
an origin in the Galactic halo for this black hole binary. 

For the thick-disk scenario, we carefully inspected the model
predictions, and although they provide better fits to the observed
abundances, they require substantial fallback (up to 5.5~\Msun) and
very efficient mixing processes between the inner layer of the
explosion and the ejecta. We thus conclude that this scenario is
unlikely. 

Metal-rich spherically symmetric models for the thin-disk scenario
were able to fairly reproduce the observed abundances, although they do
not easily provide the energy required to launch the system from
the Galactic plane to its current halo orbit. 

Finally, non-spherically symmetric models produce excellent agreement
with the observed element abundances in the secondary star without
invoking extensive fallback and mixing. In addition, asymmetric
mass ejection would naturally provide the kick to expel this binary
system from its birth place in the Galactic thin disk, which seems to
be the most plausible explanation for the origin of this halo black hole
X-ray binary.

\acknowledgments

We thank Keiichi Maeda for providing us with his aspherical explosion
models, and for helpful discussions. We are grateful to Tom Marsh for
the use of the MOLLY analysis package. The W. M. Keck Observatory is
operated as a scientific partnership among the California Institute of
Technology, the University of California, and NASA; it was made
possible by the generous financial support of the W. M. Keck Foundation. 
This work has made use of the VALD database and IRAF facilities. J.I.
acknowledges support from the EU contract MEXT-CT-2004-014265
(CIFIST). Additional funding was provided by Spanish Ministry project 
AYA2005--05149, as well as by US National Science Foundation grants
AST--0307894 and AST--0607485 to A.V.F.

\end{document}